\documentclass[10pt,journal,compsoc]{IEEEtran}

\ifCLASSOPTIONcompsoc
  \usepackage[nocompress]{cite}
\else
  \usepackage{cite}
\fi
\ifCLASSINFOpdf
\else
\fi
\usepackage{amsmath,amssymb,amsfonts,bm}
\usepackage{graphicx}
\usepackage{textcomp}
\usepackage{xcolor}
\usepackage[utf8]{inputenc}
\usepackage{tabulary}
\usepackage{booktabs}
\usepackage{subfigure}
\usepackage{cases}
\usepackage{multirow}
\usepackage{bbding}
\usepackage{caption}
\usepackage{makecell}
\usepackage[linesnumbered,ruled,vlined]{algorithm2e}
\usepackage[noend]{algpseudocode}
\usepackage{ulem}
\usepackage[]{lineno}

\newcommand{\degree}{^\circ}

\begin{document}
%
\title{AGCM-3DLF: Accelerating Atmospheric General Circulation Model via 3D Parallelization and Leap-Format}

\author{Hang~Cao,
        Liang~Yuan,~\IEEEmembership{Member,~IEEE},
        He~Zhang,
        Yunquan~Zhang,~\IEEEmembership{Senior Member,~IEEE},
        Baodong~Wu,
        Kun~Li,
        Shigang~Li,~\IEEEmembership{Member,~IEEE},
        Yongjun~Xu,~\IEEEmembership{Member,~IEEE},
        Minghua~Zhang,
        Pengqi~Lu,
        Junmin~Xiao
\IEEEcompsocitemizethanks{
  \IEEEcompsocthanksitem Hang~Cao, Kun~Li, and Pengqi Lu is with the Institute of
  Computing Technology, Chinese Academy of Sciences; and University of
  Chinese Academy of Sciences.\protect\\
  E-mail: hangcao.cs@gmail.com, \{likun,lupengqi18s\}@ict.ac.cn

  \IEEEcompsocthanksitem Liang~Yuan, Yunquan~Zhang, Yongjun~Xu, and Junmin~Xiao 
  are with the Institute of
  Computing Technology, Chinese Academy of Sciences.\protect\\
  E-mails: \{yuanliang,zyq,xyj,xiaojunmin\}@ict.ac.cn

  \IEEEcompsocthanksitem He~Zhang and Minghua~Zhang are with the Institute of
  Atmospheric Physics, Chinese Academy of Sciencess.\protect\\
  E-mails: \{zhanghe,mhzhang\}@mail.iap.ac.cn

  \IEEEcompsocthanksitem Baodong~Wu is with the Sensetime Research.\protect\\
  E-mails: wubd.cs@gmail.com
  
  \IEEEcompsocthanksitem Shigang~Li is with the Department of Computer Science, 
  ETH Zurich.\protect\\
  E-mails: shigangli.cs@gmail.com
}
 }

\markboth{Journal of \LaTeX\ Class Files,~Vol.~14, No.~8, August~2015}%
{Shell \MakeLowercase{\textit{et al.}}: Bare Advanced Demo of IEEEtran.cls for IEEE Computer Society Journals}

\IEEEtitleabstractindextext{%
\begin{abstract}
The Atmospheric General Circulation Model (AGCM) has been an important research tool in the study of climate change for decades.
As the demand for high-resolution simulation is becoming urgent, the scalability and simulation efficiency is faced with great challenges, especially for the latitude-longitude mesh-based models.
In this paper, we propose a highly scalable 3D atmospheric general circulation model based on leap-format, namely AGCM-3DLF.
Firstly, it utilizes a 3D decomposition method allowing for parallelism release in all three physical dimensions.
Then the leap-format difference computation scheme is adopted to maintain computational stability in grid updating and 
avoid additional filtering at the high latitudes. 
A novel shifting window communication algorithm is designed for parallelization of the unified model.
Furthermore, a series of optimizations are conducted to improve the effectiveness of large-scale simulations.
Experiment results in different platforms demonstrate good efficiency and scalability of the model.
AGCM-3DLF scales up to the entire CAS-Xiandao1 supercomputer (196,608 CPU cores), 
attaining the speed of 11.1 simulation-year-per-day (SYPD) at a high resolution of 25KM.
In addition, simulations conducted on the Sunway TaihuLight supercomputer exhibit a 1.06 million cores scalability with 36.1\% parallel efficiency.

\end{abstract}

}

\maketitle

\IEEEdisplaynontitleabstractindextext

%
\IEEEpeerreviewmaketitle

\IEEEraisesectionheading{\section{Introduction} \label{sec:Intro}}

In the study of climate change, the Atmospheric General Circulation Model (AGCM) has always been a critical research tool.
Given the urgent need for climate modeling at ever-finer resolutions, the numerical simulation of AGCM faces a significant challenge in terms of scalability and simulation efficiency.
The two main modules of the AGCM are the dynamical core and the physical process~\cite{zhang2013sensitivity}.
The dynamical core refers to the formulation of the atmospheric hydrodynamic equations as well as the computational algorithms to solve them, and it is one of the most time-consuming modules.

Some recently developed atmospheric models include CAM5~\cite{liu2012toward} from the National Center for Atmospheric Research (NCAR)~\cite{hurrell2013community}, ECHAM-6~\cite{popke2013climate} from the Max Planck Institute for Meteorology, HadCM~\cite{johns2003anthropogenic} from the UK Met office Hadley Center, MRI-AGCM~\cite{murakami2012future} from the Meteorological Research Institute, and IAP-AGCM~\cite{zhang2009computational} from the Institute of Atmospheric Physics, Chinese Academy of Sciences. 
Basically, there are two types of mesh that are used in these AGCMs: the quasi-uniform polygonal mesh and the equal-interval latitude-longitude mesh.
CAM-SE of NCAR adopts the first type of mesh with the spectral element dynamical core implementation and is known for its good scalability and parallel performance. CAM-FV and IAP-AGCM use the second one.
They employ a finite volume implementation and a finite-difference dynamical core, respectively.
Generally, latitude-longitude mesh-based models such as the IAP-AGCM provide benefits over quasi-uniform polygonal mesh-based dynamical cores in terms of energy conservation, computation of complex terrains and moistures, and the coupling with other climate system modules.

The IAP-AGCM4 is the latest version of the AGCM module in CAS-ESM (Chinese Academy of Sciences’ Earth System Model), which employs a finite-difference scheme with a terrain-following $\sigma$ coordinate vertically, and a latitude–longitude grid with C grid staggering in the horizontal discretization~\cite{zhang2013sensitivity}.
Despite the advantages of IAP-AGCM's dynamical core, it is still difficult to increase parallel scalability while maintaining computation stability at finer horizontal resolutions such as $0.5\degree \times 0.5\degree$ and $0.25\degree \times 0.25\degree$.
Previous work~\cite{wang2017scalable,xiao2018communication} has developed a scalable finite-difference dynamical core based on the latitude-longitude mesh using a 2D decomposition method at the resolution of 140KM and 50KM. 
The parallelism of the approach is only exploited in two dimensions (longitudinally and vertically).
Thus, the scaling limit can be easily reached.

For the latitude-longitude mesh-based models, a zonal filtering scheme is applied at the polar regions to solve the computation instability problem, also known as the pole problem~\cite{williamson2007evolution}.
Since the zonal mesh interval decreases rapidly in high latitudes~\cite{zhang2013implementation}, filtering is necessary for dampening the high-frequency effects of the shortwave and thus maintain computational stability. 
However, the high-latitude filtering brings in inevitable load imbalance along the latitudinal dimension and the MPI communication overheads will be very high.

In the original version of IAP-AGCM, a longitudinal Fast Fourier Transformation (FFT) and a Gaussian scheme are optional as the filtering module in the 2D decomposition dynamical core, both of which scale poorly as the model resolution increases up to 25KM.
For the sake of performance improvement along with the alleviation of load imbalance in large-scale simulation, we propose a highly scalable decomposition method with a newly designed parallel finite-difference computing scheme that scales well in different platforms at fine horizontal resolutions.
The key contributions of this paper are as follows:
\begin{itemize}
\item We present a 3D decomposition method that allows parallelism to be released in all three physical dimensions.
The 3D decomposition method significantly increases the parallelism and the scalability of IAP-AGCM.
\item We propose a new leap-format finite-difference computation scheme that is capable of maintaining computational stability in grid updating and avoids additional filtering at the high latitudes and polar regions.
Thus, the overall communication overhead is significantly reduced and the load imbalance is relieved.

\item  
We introduce a novel shifting window communication algorithm for parallelizing the 3D decomposed model. Several communication optimizations are utilized in addition.
Our new approach exhibits good efficiency and scalability.

\end{itemize}

This paper extends conference papers \cite{wu2018agcm3d,caoipdps20}. 
In particular, it adds: 
1) A unified integration of the 3d dynamical core and the leap-format difference computation, namely AGCM-3DLF;
2) A set of new optimizations for the efficiency of large-scale simulations, including 
decomposition prior strategy, refactoring of computation, aggregation of related variables, etc.
3) Hybrid parallelization and acceleration of stencil loops in the heterogeneous many-core platform;
and 4) Sufficient numerical tests and performance experiments for the complete AGCM model. 

The rest of this paper is organized as follows.
Section~\ref{sec:BG} describes the background. 
Section~\ref{sec:Leap} discusses the 3D decomposition method and the leap-format computation scheme along with the design of the parallell implemention.
Experimental results and analysis are presented in Section~\ref{sec:perf}. Section \ref{sec:conclu} concludes.

\section[]{Background}
\label{sec:BG}

\subsection{Dynamical Core of IAP-AGCM}\label{sec:BG.MD}

The dynamical core is one key component of an Atmospheric General Circulation Model.
The fourth generation of global AGCM developed by the Institute of Atmospheric Physics, CAS \cite{zhang2009computational}, as known as IAP-AGCM4, has been integrated into CAS-ESM to simulate the air temperature, summer precipitation, and circulations related to monsoons in the long-term atmospheric circulations and climate change \cite{xiao2012evaluation,zheng2014relationship,su2014simulating,adeniyi2019evaluation}. 

The IAP-AGCM4 employs an implicit finite-difference discretization scheme
based on the baroclinic primitive equations that utilize a latitude-longitude grid with C grid staggering in the horizontal discretization.
The discretized governing equations can be written as follows after subtracting standard stratification, IAP transform, and terrain-following vertical coordinate:
\renewcommand{\arraystretch}{1.8}
\begin{equation}
    \left\{
    \begin{array}{lll}
        \frac{\partial U}{\partial t}=-\sum_{m=1}^3{\alpha ^*L_m\left( U \right)}-\beta ^*P_{\lambda}-\gamma ^*f^*V		\\
        \frac{\partial V}{\partial t}=-\sum_{m=1}^3{\alpha ^*L_m\left( V \right)}-\beta ^*P_{\theta}+\gamma ^*f^*U		\\
        \frac{\partial \Phi}{\partial t}=-\sum_{m=1}^3{\alpha ^*L_m\left( \Phi \right)}+\left( 1-\delta _p \right) 		\\
        \quad \quad \ \; \cdot \left[ b\left( 1+\delta _c \right) +\delta \cdot \kappa \Phi /P \right] \cdot \beta ^*\widetilde{\Omega }	\\
        \frac{\partial}{\partial t}\left( p_{sa}^{'}/p_0 \right) =-\beta ^*\widetilde{P}\left( W \right) +\kappa ^*D_{sa}/P_0		\\
    \end{array}
    \right. \label{eq:ctrdf1}
\end{equation}
where the partial derivatives $U$, $V$, $\varPhi$, $p_{sa}^{'}$ and $\phi ^{'}$, $W$ are the forecast variables and prognostic variables, respectively, 
representing the calculations of different variables’ tendencies.


Generally, the large-scale motion in the dynamical core of the atmosphere is divided into two parts: the advection process and the adaption process~\cite{zhh2011mutual}, which together account for the majority of the overall execution cost. For the sake of briefness and energy conservation, the governing equations (\ref{eq:ctrdf1}) can be written as follows:
\begin{equation}
    [ \frac{\partial F}{\partial t} ] _{x,y,z}  =-\left[ L_F \right] _{x,y,z}+\left[ A_F \right] _{x,y,z} 
    \label{eq:timeint}
\end{equation}
where $x$, $y$, $z$ denote the longitudinal, latitudinal, and vertical
index of the 3D mesh, and $F$ represents the forecast variables ($U,V,\varPhi ,p_{sa}^{'}$).
The functions $L$ and $A$ correspond to the adaption process and the advection process, respectively.


\subsection{Domain Decomposition and Parallelism}
\label{sec:BG.Decomp}

\begin{table}[bp]
    \caption{The numbers of mesh grids in different horizontal resolutions}
    \begin{center}
    \begin{tabular}{m{3cm}<{\centering} m{1.3cm}<{\centering} m{1.3cm}<{\centering} m{1.3cm}<{\centering}}
    \toprule 
    \textbf{Horizontal resolution} & \textbf{$\bm {N_x}$ }& 
    \textbf{$ \bm {N_y}$ }  & \textbf{$ \bm {N_z}$ } \\
    \midrule 
    $1.4\degree\times1.4\degree$&   256   &   128 &  30  \\ 
    $0.5\degree\times0.5\degree$&   768   &   361 &  30  \\ 
    $0.25\degree\times0.25\degree$& 1,152 &   768 &  30 \\ 
    \bottomrule
    \end{tabular}
    \label{tab:num}
    \end{center}
\end{table}

The original dynamical core of IAP-AGCM4 adopts a 2D decomposition scheme, where only the latitudinal ($Y$) and vertical ($Z$) dimensions are split to provide parallelism.
Meanwhile, the physical process is decomposed along the longitudinal ($X$) and latitudinal ($Y$) dimensions. 
That's because the physical parameterizations utilize a vertical single-column model, where the computations of each mesh grid are irrelevant to the neighbor horizontal points. In addition, the dynamical core and the physical process need to be coupled to each other for data transmission, so the number of processes of the two modules is typically equal.

With this Cartesian topology, the dynamical core's parallelism is restricted to a great extent since the $X$ dimension usually contains the greatest number of grid points. This limitation becomes severer when performing large-scale simulations for high-resolution models.
Conventionally, the maximum number of processes IAP-AGCM can scale up to is described as follows:
\begin{align}    
    P_{xyz} &= P_y\times \min (P_x, P_z) \notag \\
            &= \lfloor \frac{N_y}{n_y} \rfloor \times \min ( \lfloor \frac{N_x}{n_x} \rfloor ,\lfloor \frac{N_z}{n_z} \rfloor ) 
\label{eq:2dscale}
\end{align}
where $N_x,N_y,N_z$ are the total numbers of mesh points in corresponding dimensions, $n_x,n_y,n_z$ are the minimum numbers of points for one process, and $P_{xyz}$ is the overall number of processes for the model.
The numbers of mesh grids in different resolutions are shown in Table~\ref{tab:num}.
Using the resolution $1.4\degree\times1.4\degree$ as an example, and assuming $n_x,n_y,n_z$ to be 2, 2 and 1, respectively, $P_{xyz}$ equals $\lfloor {128 / 2} \rfloor \times \min ( \lfloor {256 / 2} \rfloor ,\lfloor  {30 / 1} \rfloor ) $, i.e, the scalability of $1.4\degree\times1.4\degree$ model is limited to 1,920 processes.

With the imperious need for high-resolution numerical simulation of AGCM, the original 2D decomposed model is no longer effective enough to take advantage of the rapidly growing computing resources of current supercomputers. As also is the case of the NCAR finite-volume dynamical core based on the latitude-longitude mesh, which can only scale up to 1,6000 MPI processes~\cite{dennis2012cam,evans2013amip} at the resolution of 0.25$\degree$.



\subsection{Filtering and Load-balancing}
\label{sec:BG.Filt}
For the latitude-longitude mesh-based models, the finite-difference discretization results in unequal longitudinal distances.
Due to the convergence of meridians toward the polar regions, the physical distance between adjacent mesh points would decrease rapidly~\cite{randall2002climate,wang2004design}. 
To satisfy the Courant-Friedrichs-Lewy (C.F.L) condition~\cite{courant1928partiellen} for the PDE solver's convergence, the model should meet the condition
$$\varDelta t\leqslant \frac{\varDelta x}{U}$$ 
where $\Delta t$ and $\Delta x$ are the time step and the space interval, and $U$ is the maximum characteristic velocity, respectively. 

To maintain computational stability while allowing for a larger time step and lower computation expense, a high-frequency filtering module is adopted poleward of $\pm 70\degree$.
In the former version of IAP-AGCM, an FFT scheme and a Gaussian scheme are optional as the filtering modules to dump out the short-wave modes at the polar regions, and both of the two schemes have their own merits and shortcomings.
The FFT scheme, which originates from the widely-used FFT99~\cite{temperton1988self}, is performed sequentially at each latitudinal ($X$) circle.
\begin{table}[bp]
    \caption{The runtime ratio of filtering modules in the whole dynamical core}
     \label{tab1}
    \begin{center}
     \setlength{\tabcolsep}{1.2mm}{ \begin{tabular}[0.5\textwidth]{lccccccccc}    
        \toprule
    \textbf{Num of procs} & 
    \textbf{128} & \textbf{256} & \textbf{512} & \textbf{1024} & \textbf{2048} & \textbf{4096} & \textbf{8192}  \\
        \midrule
    FFT & 28.8\% & 17.3\% & 6.9\% & 8.3\% & N/A & N/A & N/A  \\
    Gaussian & 58.1\% & 40.4\% & 21.3\% & 27.2\% & N/A & N/A & N/A  \\
    Gaussian (3D) & 69.6\% & 81.9\% & 90.7\% & 94.6\% & 96.1\% & 95.3\% & 91.1\%  \\
        \bottomrule
    \end{tabular}}
    \end{center}
\end{table}
Despite the already demonstrated efficiency and maturity~\cite{temperton1992generalized} for low-resolution models, the FFT scheme might fail to meet the needs to further improve the parallelism for finer resolutions. Also, the frequently called collective communications in parallel FFT is one key factor for the 2D domain decomposition strategy.
The Gaussian filtering~\cite{ito2000gaussian} is a self-adaptive filtering scheme, which can obtain equivalent effectiveness to the FFT scheme while avoiding the costly all-to-all communication~\cite{wu2018agcm3d} along the $X$ dimension. Whereas, the overall computational overhead Gaussian scheme brings in increases by a large amount. The reason is that it takes multiple filtering calls to ensure the stability of the model, and the number of calls for each forecast variable is related to the current latitude, which could be as high as 25 at each iteration.

\begin{figure}[tbp]
    \centerline{\includegraphics[width=3.5in]{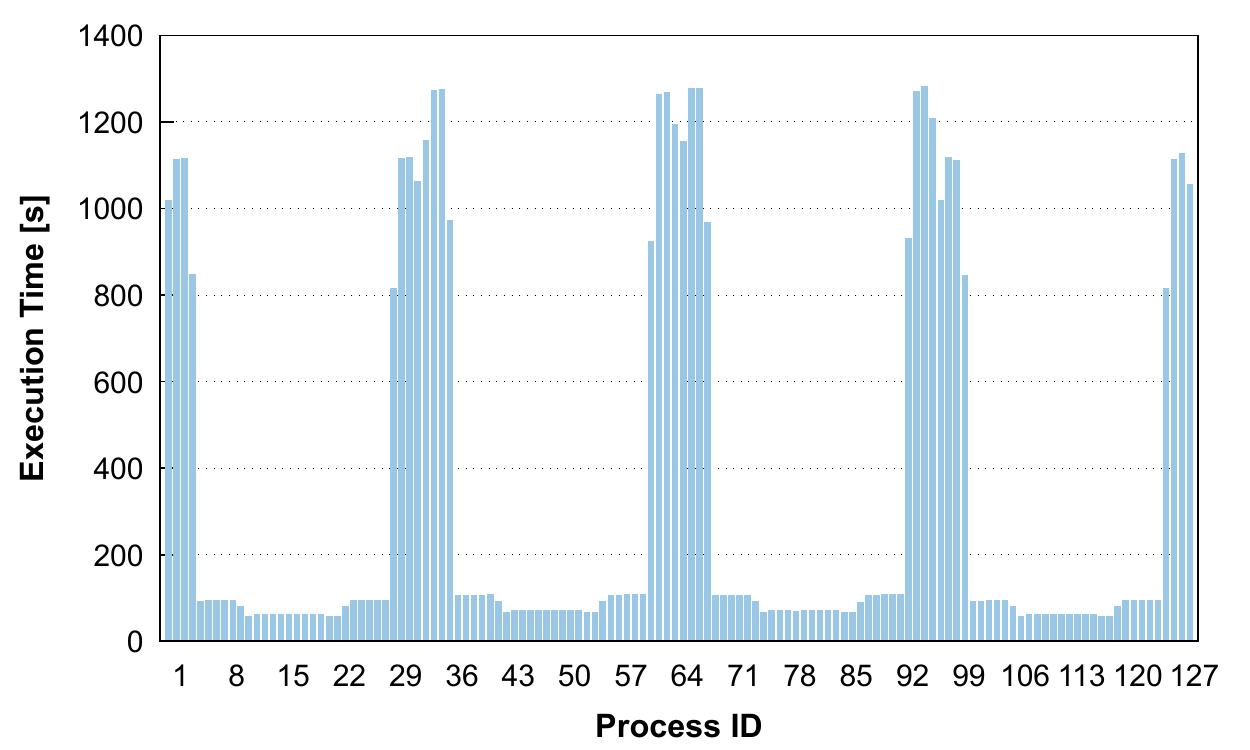}}
    \caption{Runtime of the filtering module in different processes indicates the load imbalance.}
    \label{Filt_overhead}
 \end{figure}

Except for the absolute costs of the filtering schemes, the load-balancing between different processes is not to be ignored. 
Along with the further decrease of the longitudinal ($X$) mesh size in the high-resolution model, the overhead of the filtering module in each latitudinal circle will increase rapidly, resulting in a severe load imbalance. 
The filtering costs of 128 processes with a $32 \times 4$  (Y $\times$ Z)  decomposed simulation are shown in Fig.~\ref{Filt_overhead}. 
The computation costs for high-latitude processes are much higher. As a result of the load imbalance of filtering in the $Y$ dimension, there is significant difficulty in further parallelizing the model and providing better scaling in large-scale computing systems.

\section{3D Domain Decomposition and Leap-\\
Format Difference Scheme}
\label{sec:3D-leap}

In this section, we will introduce our new 3D decomposition method, and the leap-format approach to the high-latitude and polar problems. 
Firstly, we propose the 3D domain decomposition in Section~\ref{sec:3D_domain}.
Then the leap-format finite-difference computation scheme is introduced in Section~\ref{sec:Leap}.
Section~\ref{sec:para-LF} presents the 3D implementation with leap-format computation.
\begin{figure}[htbp]
  \centering
    \subfigure[2D domain decomposition.]{
      \label{2d-decomp} 
      \includegraphics[width=1.6in]{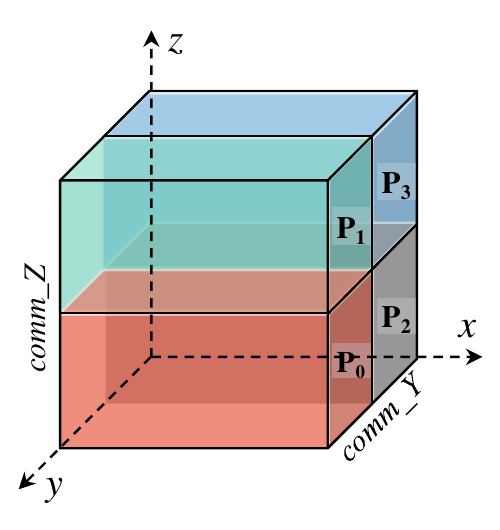}}
    \subfigure[3D Domain decomposition.]{
      \label{3d-decomp} 
      \includegraphics[width=1.6in]{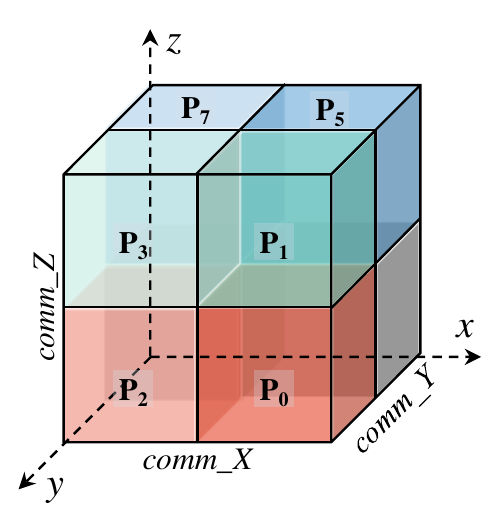}}
    \caption{From the original 2d domain decomposition to the new 3D decomposition scheme.}
  \label{Domain}
\end{figure}
Furthermore, a series of efficient optimizations are utilized in Section~\ref{sec:Opt}.
Finally, we present the hybrid parallelization and acceleration of stencil loops in the heterogeneous many-core platform TaihuLight in Section~\ref{sec:SW}.
\subsection{3D Domain Decomposition}
\label{sec:3D_domain}

\renewcommand{\thempfootnote}{\arabic{mpfootnote}}
\begin{table}[bp]
  \caption{The comparison between 2D and 3D decompositions at the horizontal resolution of $0.25\degree\times0.25\degree$}
  \begin{center}
  \begin{minipage}{.49\textwidth}
  \begin{tabular}{ m{3.0cm}<{\centering} m{2.3cm}<{\centering} m{2.3cm}<{\centering} }
  \toprule 
  \textbf{Comparison items} & \textbf{\textit{2D}}& \textbf{\textit{3D}} \\
  \midrule 
  Number of mesh points $\left( N_x\times N_y\times N_z \right)$   & \multicolumn{2}{c}{$1152\times768\times30$} \\ 
  Theoretical parallelism& $1 \times P_y \times P_z$  &  $P_x \times P_y \times P_z$   \\  
  Minimum communication volume along X per core\footnote{Excluding the filtering overhead}
                                                &$0$& $(\alpha+\beta\times \frac{N_z}{P_{z}})\times \frac{N_y}{P_{y}}$ \\ 
  Minimum communication volume along Y per core & $(\alpha+\beta\times\frac{N_z}{P_{z}})\times N_x$ &
                                                $(\alpha+\beta\times \frac{N_z}{P_{z}})\times \frac{N_x}{P_{x}}$ \\ 
  Minimum communication volume along Z per core & $\beta\times\frac{N_y}{P_{y}}\times N_x$&
                                                $\beta\times\frac{N_y}{P_{y}}\times \frac{N_x}{P_{x}}$ \\ 
  Minimum collective communication volume along Z per core\footnote{Only when $P_z > 0$} & $\frac{N_y}{P_{y}}         
                                                \times\frac{N_z}{P_{z}}\times N_x$&
                                                $\frac{N_y}{P_{y}}\times\frac{N_z}{P_{z}}\times\frac{N_x}{P_{x}}$\\ 
  \bottomrule
  \end{tabular}
  \end{minipage}
  \label{tab:2dN3dcomm}
  \end{center}
\end{table}
As previously mentioned, the original 2D domain decomposition hinders parallel scalability when the $X$ dimension remains serialized.
The new three-dimensional decomposition method releases parallelism in all three physical dimensions. 
Following that, the primary forecast and prognostic variable arrays are then split and assigned into corresponding processes by the newly introduced $X$ communicator.
Assume there are $N_x$, $N_y$, $N_z$ mesh points and $P_{x}$, $P_{y}$, $P_{z}$ processes for each of the $X$, $Y$ and $Z$ dimensions. 
For the original 2D domain decomposition (Y-Z) as shown in Fig.~\ref{2d-decomp}, the global communicator is split into two sub-communicators, i.e., $comm\_Y$ and $comm\_Z$, and each process is in charge of $\left( N_x\times N_y\times N_z \right) /\left( P_x\times P_y \right) $ mesh points. 
The new 3D domain decomposition method provides a new sub-communicator ($comm\_X$) along the longitudinal dimension, as exhibited in Fig.~\ref{3d-decomp}. And the maximum number of processes the 3D dynamical core can scale to is as follows:
\begin{linenomath}
  \begin{align}   
  P_{xyz} &= P_x\times P_y\times P_z \notag \\
          &= \lfloor \frac{N_x}{n_x} \rfloor \times \lfloor \frac{N_y}{n_y} \rfloor \times \lfloor \frac{N_z}{n_z} \rfloor 
  \end{align}
\end{linenomath}
where $n_x,n_y,n_z$ are the minimum numbers of mesh points for one process, and $P_{xyz}$ is the overall number of processes for the model.
Obviously, the parallelism is increased by $P_{x}$ fold compared to Equation~(\ref{eq:2dscale}).

Except for the increased parallelism, there are a few advantages that the 3D domain decomposition method brings in aspects of communication overhead.
Firstly, we take the finest resolution $0.25\degree\times0.25\degree$ in IAP-AGCM so far as an example. 
The comparison of communication overhead between the 2D and 3D domain decomposition is shown in Table~\ref{tab:2dN3dcomm}. 
As can be seen, the minimum per core point-to-point communication volume in both Y and Z dimensions is reduced by $P_{x}$ times. 
Note that $\alpha$ and $\beta$ represent the amount of 2D variables and 3D variables, respectively. 
Also, the time-consuming collective communication along the Z dimension is reduced to $\frac{N_y}{P_{y}}\times\frac{N_z}{P_{z}}\times\frac{N_x}{P_{x}}\times sizeof(DOUBLE)$ per core (when $P_z > 0$).
Despite the addition of point-to-point communication for stencil computations along the X dimension, 
the overall communication overhead is significantly decreased using the new 3D domain decomposition.
Secondly, during the overall iteration of the IAP-AGCM model, 
the 3D dynamical core needs to couple with the physical process at certain intervals, 
which practically executes the data transfers by means of memory copies (intra-process) or MPI communication (inter-process). 
The implementation of D-P coupling in fact requires the two main modules to employ an equal number of processes. 
Thus, the parallelism of physical processes, though known as highly scalable, is limited to the scalability of the 2d dynamical core. 
With the new 3D domain decomposition method, the parallelism of IAP-AGCM can be further released.

\subsection[]{Leap-Format Difference Scheme} 
\label{sec:Leap}

During the iteration of the dynamical core, the filtering module is required to alleviate the serious short wave impact for the latitude-longitude mesh-based AGCM model.
In detail, the conventional filtering scheme often distinguishes different latitude zones, as shown in Table~\ref{tab:Filt_scheme}. 
For the lowed latitude zones, i.e, $|\varphi| < 70\degree$, a simplified filter ($0\degree < |\varphi|<38\degree$) or a 3-point recursive operator ($38\degree \leqslant|\varphi| \leqslant 70\degree$)\cite{charney1990numerical} is adopted to relieve the impact of double mesh spacing waves.
The stronger FFT filtering and adaptive Gaussian filtering focus on dealing with variables at high latitudes.


\begin{figure}[htbp]
  \centerline{\includegraphics[width=2.2in]{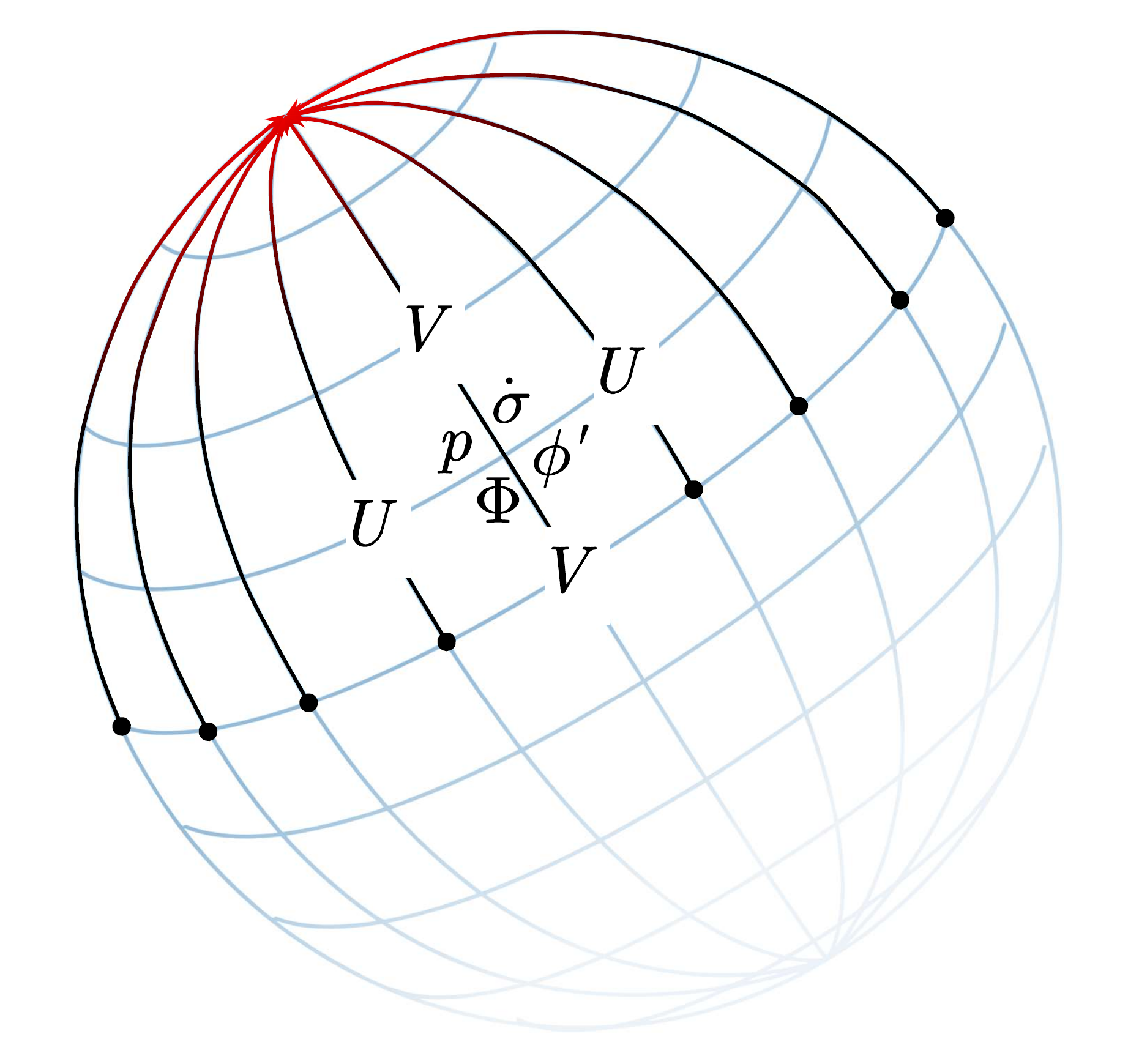}}
  \caption{Distribution of variables in C grid.}
  \label{Cgrid}
\end{figure}

We observe that with the utilization of 3D Domain decomposition in IAP-AGCM, the FFT filtering will incur $logP_x$ times of all-to-all communication opeartions along the $X$ dimension and the communication volume of $\frac{N_y}{P_{y}}\times\frac{N_z}{P_{z}}\times N_x$ for each process. 
Considering the huge overhead of all-to-all collective communication, the Gassian filtering scheme that whereas involves a large amount of point-to-point communication calls seems to be the preferred but not the perfect choice in the dynamical core.
\begin{table}[bp]
  \caption{Conventional Filtering scheme}
  \begin{center}
    \renewcommand\arraystretch{2}
  \begin{tabular}{ m{5.5cm}<{\centering} m{2.3cm}<{\centering} }
  \toprule 
  \textbf{Filtering Scheme}& \textbf{Latitudinal band} \\
  \midrule 
  $F_{x,y}\pm \frac{2}{I}\sum_{x=1}^{\frac{I}{2}-1}{\left( F_{2x,y}-F_{2x+1,y} \right)}$ 
        & $0\degree < |\varphi|<38\degree$ \\ 
  $F_{x,y}^{n-1}+r_j\left[ F_{x-1,y}^{n-1}-2F_{x,y}^{n-1}+F_{x+1,y}^{n-1} \right]$ 
        & $38\degree \leqslant|\varphi| \leqslant 70\degree$ \\  
  $\sum_{n=-2K}^{2K}F_{(x+n),y}*W_{x,y;x+n}$ (FFT)  
        & \multirow{2}{*}{$70\degree <  |\varphi| < 90\degree$} \\ 
  $\sum_{n=-\varepsilon }^{\varepsilon }F_{(x+n),y}*W_{x,y;x+n}$ (Gaussian) 
        & \\ 
  \bottomrule
  \end{tabular}
  \label{tab:Filt_scheme}
  \end{center}
\end{table}

Another observation is that the finite-difference computation at high latitudes offers a complimentary property of being adjustable in terms of format design.
Particularly, larger mesh spacing in polar regions won't destroy the conservation of the finite-difference format along the $X$ dimension.
Our approach aims to avoid the multiple calls of additional filtering demands at high latitudes by reforming the finite-difference scheme and incorporating the filtering function directly with the difference calculation.
Considering that simpler filters are adopted at lower latitudes, the primary cause of load imbalance is the high latitudinal filtering.
Thus our new leap-format scheme is expected to significantly boost performance.

In most AGCMs including IAP-AGCM, the dynamical core mainly consists of two components: the advection process and the adaption process. 
IAP-AGCM employs a latitude-longitude grid for both processes, with Arakawa's C grid staggering in the horizontal discretization.
The variables are calculated in three dimensions, namely longitudinal ($X$), latitudinal ($Y$), and vertical ($Z$) dimensions.
The distribution of the primary forecast and prognostic variables is illustrated in Fig.~\ref{Cgrid}.
In the Z dimension, each forecast variable or prognostic variable's vertical distribution is placed on the integer layer or semi-integer layer.
The C grid discretization staggers in the horizontal ($X$ and $Y$) dimensions. 
Generally, the forecast variable zonal wind ($U$) is set at $(x+\frac{1}{2},y,z)$, which is located in the semi-integer layer along the $X$ dimension and the integer layer along the $Y$ dimension. 
The meridional wind $V$ is set at $(x,y+\frac{1}{2},z)$ 
while other forecast variables are set at $(x,y,z)$. 

\begin{figure}[tbp]
  \centerline{\includegraphics[width=3.2in]{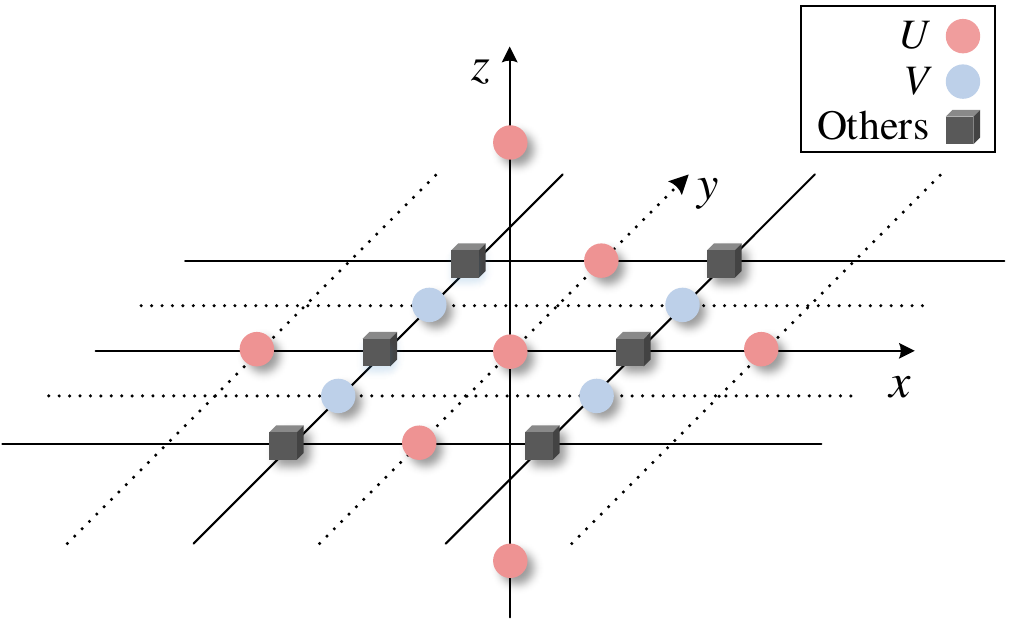}}
  \caption{Stencil computation for 3D variables.}
  \label{stencil}
\end{figure}

The calculations of the main forecast variables are typical 3D7P star stencil computations~\cite{yuan2019tessellating} distributed in staggered grids.
We take the $U$ and $V$ as the simplified examples of the stencil computations again, as shown in Fig.~\ref{stencil}. 
The integer layer and the semi-integer layer along the $X$ dimension are denoted as the solid and dotted lines, respectively.
When focusing on $U$, the three-dimensional central difference form is as follows: 
\begin{equation} 
    \left( \frac{\partial U}{\text{a}\sin\theta \partial \lambda} \right) _{x,y,z}=\frac{U_{x+\frac{1}{2},y,z}-U_{x-\frac{1}{2},y,z}}{\text{a}\sin\theta _y\varDelta \lambda} 
    \label{eq:central_init} 
\end{equation} 
where $\theta$ denotes the colatitude ($90\degree - \varphi$) of the current grid point, $\varDelta \lambda$ is the zonal grid spacing, $a$ is the Earth radius and the subscripts $x$, $y$ and $z$ denote the index of the longitudinal, latitudinal, and vertical direction, respectively. 

To obviate the restrictions caused by the filtering and the standard finite-difference format,
we propose a new leap-format finite-difference scheme. 
The fundamental technique is to adjust the mesh interval at high latitudes. 
For an exact high latitude, the zonal grid spacing $\varDelta \lambda$ utilized in the center difference as described in Equation~(\ref{eq:central_init}) is increased to a larger size. 
Thus, the subscripts can be generalized to $U_{x+N_{leap}/2,y,z}$ and $U_{x-N_{leap}/2,y,z}$, where $N_{leap}$ denotes the extended grid number of the new leapt central difference of $U$ along the $X$ direction. 
Accordingly, the grid size slids from $\Delta x$ to $\Delta x *N_{leap}$. 
The following is the new leap-format central difference form of $U$:
\begin{equation}
        \left( \frac{\partial U}{\text{a}\sin\theta  \partial \lambda} \right) _{x,y,z}=\frac{U_{x+N_{leap}/2,y,z}-U_{x-N_{leap}/2,y,z}}{\text{a}\sin\theta  _y\varDelta \lambda*N_{leap}}
    \label{eq:central_leap1}
\end{equation}

\begin{figure}[htbp]
    \centering
      \subfigure[Original central difference scheme.]{
        \label{central_leap1} 
        \includegraphics[width=3.3in]{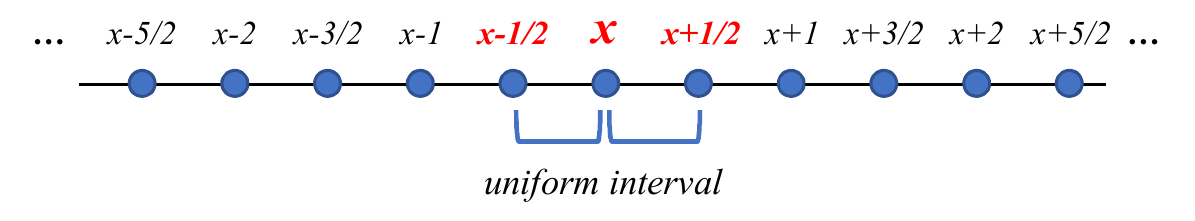}}
      \quad
      \subfigure[New central difference scheme with leap-format.]{
        \label{central_leap2} 
        \includegraphics[width=3.4in]{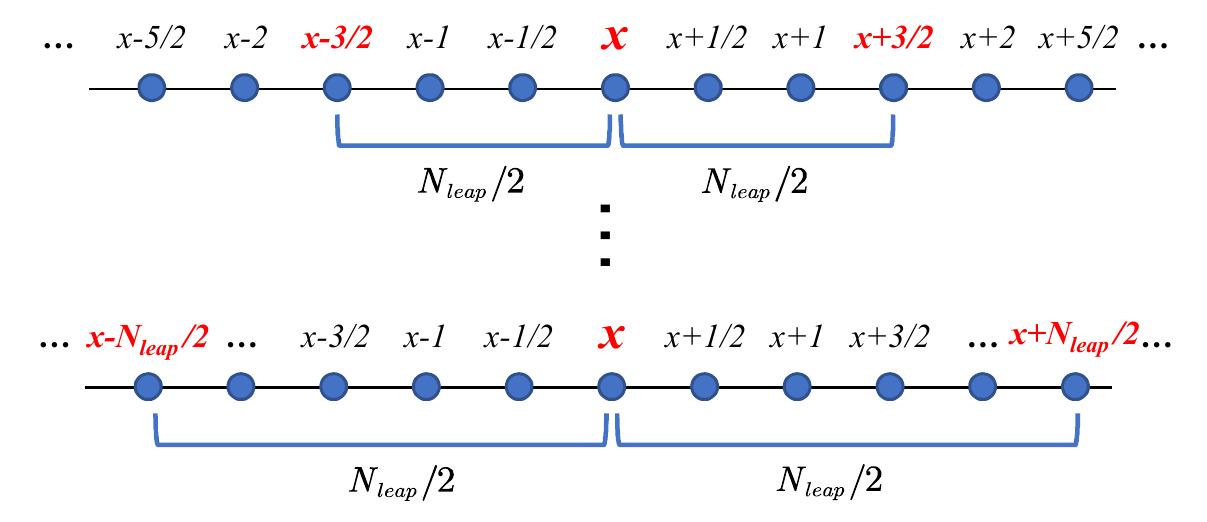}}
      \caption{Transformation of central difference scheme from fixed interval to leap-format.}
    \label{central_leap}
\end{figure}

Note that when the number of leaped points equals 1, Equation~(\ref{eq:central_leap1}) degrades into the general central difference.
And the difference terms and grid spacing do not vary with the change of grid position in other dimensions ($Y$ and $Z$) since the filter is only performed along the zonal circle. 
Based on the different latitudes for distinct grid points, the $N_{leap}$ adjusts to corresponding integer values.
Fig.~\ref{central_leap1} and Fig.~\ref{central_leap2} show the original difference scheme with a uniform interval, 
and the new central difference scheme with different leap intervals, respectively. 
Recall that the zonal wind $U$ is located in the semi-integer points in the $X$ dimension, 
thus $N_{leap}$ must be odd integers.

\begin{table}[bp]
    \caption{The zonal grid size in different resolutions}
    \begin{center}
    \begin{tabular}{m{2.0cm}<{\centering} m{1.5cm}<{\centering} m{1.5cm}<{\centering} m{1.5cm}<{\centering}}
    \toprule 
    \textbf{Horizontal Resolution} & \textbf{$\bm {\Delta  x_{\rm equator}}$ (km)}& \textbf{$\bm {\Delta x_{\rm poles}}$ (km) of u} & \textbf{$\bm {\Delta x_{\rm poles}}$ (km) of v} \\
    \midrule 
    $2\degree\times2\degree$&       222.4&   7.8&   3.9\\ 
    $1.4\degree\times1.4\degree$&   155.7&   3.8&   1.9   \\ 
    $1\degree\times1\degree$&       111.2&   1.9&   1.0    \\ 
    $0.5\degree\times0.5\degree$&   55.6&    0.5&   0.25\\ 
    $0.25\degree\times0.25\degree$& 27.8&    0.13&  0.06\\ 
    \bottomrule
    \end{tabular}
    \label{tab2}
    \end{center}
\end{table}

One major issue of the new difference scheme is that as the colatitude $\theta$ decreases, the zonal grid size ($\varDelta x = \text{a}\sin\theta\varDelta\lambda$) shrinks rapidly.    
Table~\ref{tab2} lists the zonal grid sizes $\Delta x$ for $U$ and $V$
on the equator and at the poles in different horizontal resolutions. 
A point of note is that the interval of $V$ is approximately half of that of $U$ at the poles since $\sin\theta\approx\theta$ for closest latitude circles to the poles and $V$ is located at the semi-integer layer along the $Y$ dimension, as shown in Figure~\ref{Cgrid} and Figure~\ref{stencil}. 
Take the horizontal resolution of $1.4\degree\times1.4\degree$ for example, the longitudinal physical distance between adjacent mesh points is approximately 155.7 km at the equator, while 3.8 km at the poles.
For finer resolutions, the difference of grid size between the polar regions and low latitude regions will be magnified as the horizontal girds' density increases.
As an example, for the meridional wind $V$, 
the ratio of the interval at poles to that on the equator is $222.4/3.9\approx 58$ at the resolution $2\degree$ while it increases to $27.8/0.06\approx 463$ at the resolution $0.25\degree$.

Another key technique is that we make the value of
the leap interval $N_{leap}$ adaptively adjust with the latitude's changing. 
In detail, the interval size at mid-latitude $45 \degree$ is chosen as a reference value and each interval in different high latitudes self-adjusts to an equivalent physical size with the reference.
The zonal distance of the mesh interval in the spherical coordinate system can be calculated by $2*a*\arcsin \left( \cos \alpha\times \sin res \right)$, where $a$ is the radius of earth,  $\alpha$ represents the current latitude, and $res$ is the difference of longitudes (resolution in $X$ dimension). 
Hence, $N_{leap}$ is defined by the ratio of the referenced threshold ($45 \degree$) to the mesh size of the current colatitude $\theta_y$.

\begin{equation}
        N_{leap}=\lceil \frac{\arcsin \left( \cos 45\degree\times \sin res \right)}{\arcsin \left( \cos (90\degree -\theta_y)\times \sin res \right)}\rceil
    \label{eq:central_leap2}
\end{equation}

In Fig. \ref{leap_points}, we exhibit the $N_{leap}$ values 
for different horizontal resolutions ($0.5\degree\times0.5\degree$, $0.25\degree\times0.25\degree$). 
The number of leap points can reach as high as 41 or 82 for $0.5\degree$, and even 87 or 173 for $0.25\degree$, when the colatitude approaches zero. 
Note that 
when dealing with difference terms other than the central difference, the physical interval adjusting scheme should be variant. 

From the perspective of atmospheric sciences, with the equal physical size to the referred low latitude for high latitudes,  the model's simulation allows for a larger time step,
which plays an equivalent part as the filtering modules. 
To put it another way, no extra filters are needed at high latitudes ($|\varphi|>70\degree$). 
Note that the inexpensive simple filters for lower latitudes ($ 0\degree \leqslant|\varphi| \leqslant 70\degree$) remain the same.
As a result, the implementation of the leap-format difference scheme can reduce the overall runtime of the dynamical core and facilitate load balancing in the meantime.

\subsection[]{3D Implementation with Leap-Format Computation}
\label{sec:para-LF}


 
\begin{figure}[tbp]
  \centerline{\includegraphics[width=3.2in]{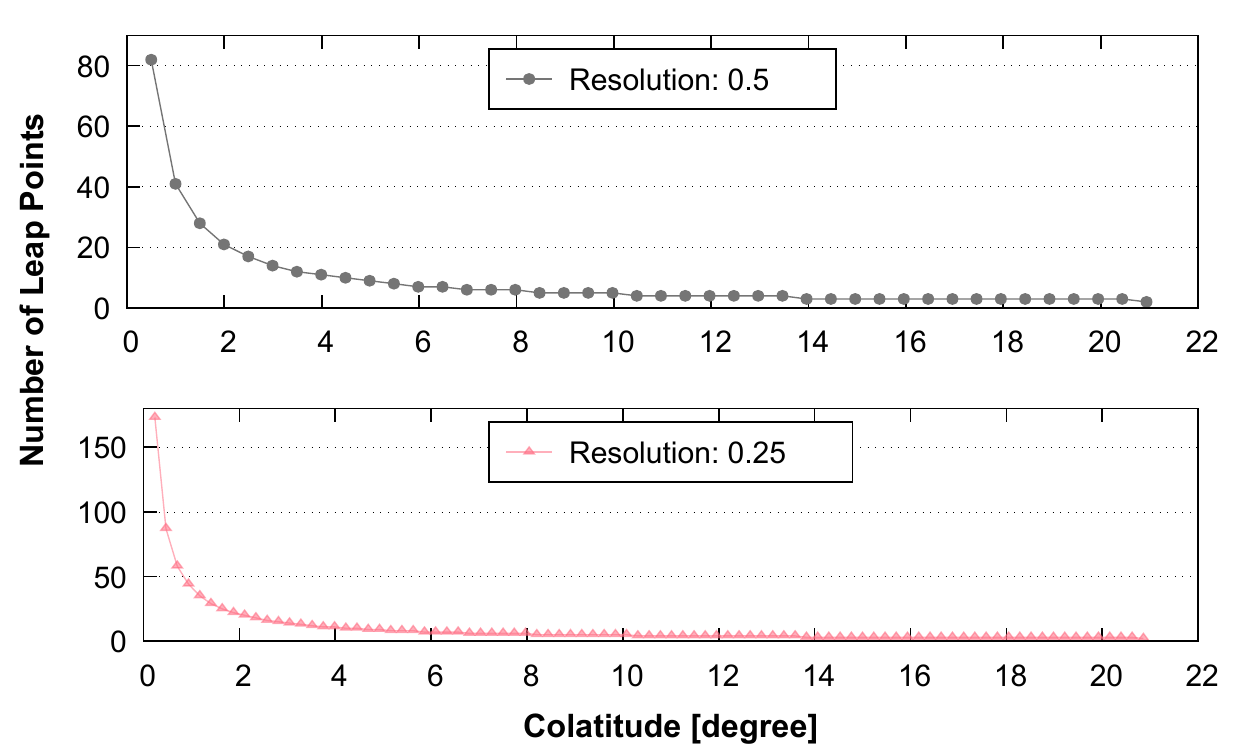}}
  \caption{Numbers of leap points in various colatitudes for different resolutions.}
  \label{leap_points}
\end{figure}

For the purpose of parallelizing the leap-format computation in the 3D decomposed AGCM, the concrete values of the leap points, namely $N_{leap}(j)$ for each latitude should be considered according to Fig. \ref{leap_points}. 
When adopting the 3D decomposition, a new $X$ sub-communicator is brought in along the $X$ dimension, i.e. the latitudinal circle direction. 
With such a wide range of leap points, the original neighbor communication between ghost regions along the $X$ sub-domain fails to meet the requirements of the new leap-format difference computation at high latitudes.

\begin{figure*}[htbp]
  \centerline{\includegraphics[width=4.7in]{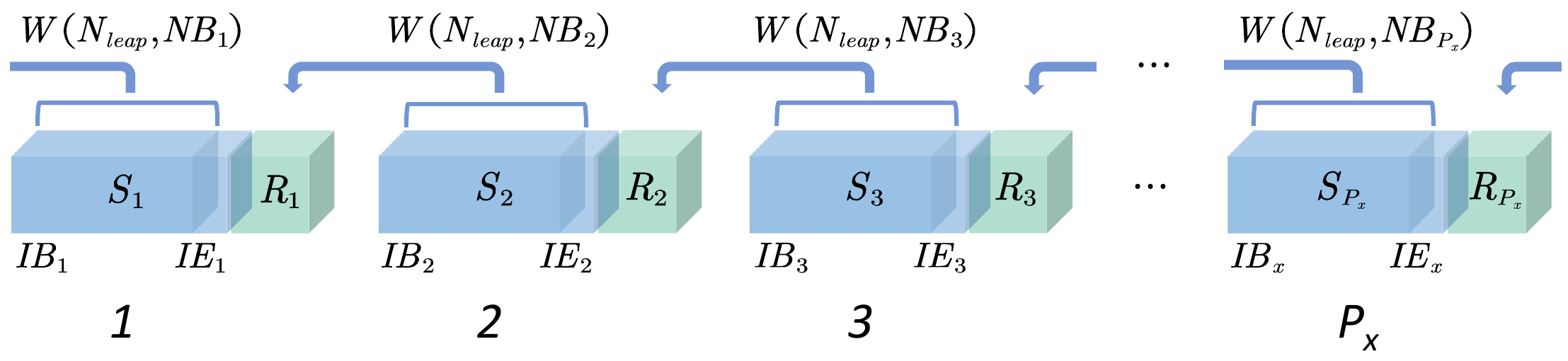}}
  \caption{Shifting leap-format communication for neighbor processes.}
  \label{leap_comm2}
\end{figure*}

\begin{figure*}[htbp]
  \centerline{\includegraphics[width=6in]{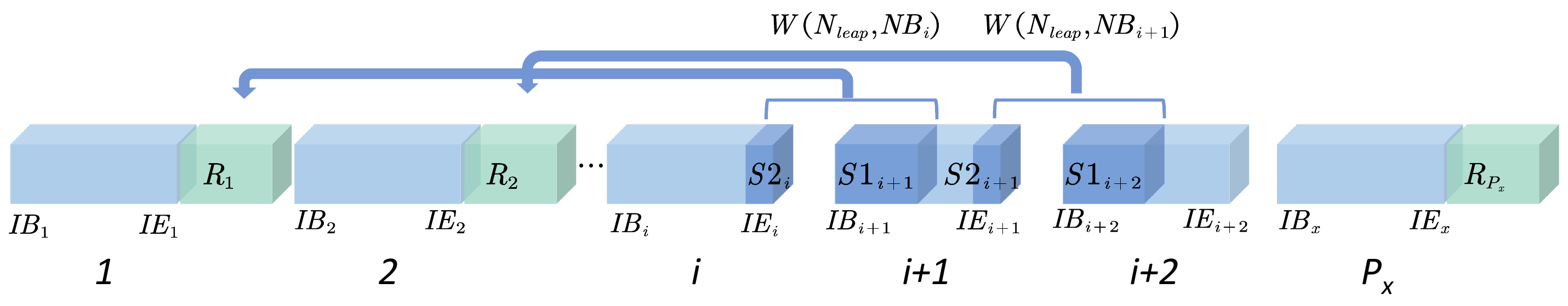}}
  \caption{Shifting leap-format communication for crossed processes.}
  \label{leap_comm3}
\end{figure*}
In different cases of the leap-format difference computations at high latitudes, various point-to-point communication calls are required to send the corresponding values from the current process to the relevant process. 
\begin{table}[htbp]
  \caption{Leap-format communication scheme}
  \begin{center}
  \begin{tabular}{m{3.5cm} m{2cm}<{\centering} m{1.5cm}<{\centering}}
  \toprule 
  \textbf{Number of leap points} & \textbf{Communication volume along $X$}& \textbf{Participant processes} \\
  \midrule 
  $1<N_{leap}\leqslant NB_{i+1}$&   $N_{leap}$&   Neighbors\\ 
  $N_{leap}> NB_{i+1}$&   $NB_{i}$&    Remote \& Crossed\\ 
  \bottomrule
  \end{tabular}
  \label{tab:leap_comm}
  \end{center}
\end{table}
Recall that $N_x$, $N_y$ and $N_z$ are the amount of mesh points along three dimensions, and 
$P_x$, $P_y$ and $P_z$ are the numbers of processes assigned in three sub-communicators. 
Based on the grid partition of the $X$ sub-communicator, the variables are separated into local arrays in different processes along the zonal($X$) direction.

The data size that each process $i$ ($ i \in \{ {1, 2, ..., P_x} \}$) possesses along the $X$ direction is referred to as $NB_i$. 
Note that 
$NB_i$ should be rounded up or down to an integer value when $N_x$ is not a multiple of $P_x$.
To apply the 3D domain decomposition method to the new leap-format finite-difference scheme, 
we introduce a shifting leap-format communication algorithm.
The core concept is that we locate the process rank and determine the size of the required data, ie, the communication window.
Assume that the starting and ending indexes of the local array for the current process are $IB_i$ and $IE_i$, respectively.
By definition, the start of a communication window can be simply calculated by $IE_i + N_{leap}/2$. 
Considering that the window size $W(N_{leap},NB_i)$ is determined by two main factors, $N_{leap}$ and $NB_i$,
we further classify the communication scenarios into two cases based on whether the point-to-point communication is solely related to the neighbor processes.
For simplicity, we only study the communication direction of sending to the left and receiving from the right.
The opposite communication direction is similar. 
As shown in Table \ref{tab:leap_comm}, the communication volume and pattern may differ when the number of leap points lies in different zones.



In Fig. \ref{leap_comm2}, we illustrate the neighbor communication case, 
where $S_i$ ($i \in \{ 1, 2, ..., P_x\} $) denotes the send buffer of the current process, 
and $R_i$ ($i \in \{ 1, 2, ..., P_x\} $) represents the receiving buffer. 
In this case, it takes one send operation and one receive operation with a data size of $N_{leap}$ for each process to fulfill the requirements of leap-format computations.


Fig. \ref{leap_comm3} exhibits the other case where the current process necessitates data from at least one remote process according to the shifting window's position. 
In this case, The data size of receiving buffers remains $N_{leap}$, and the data size of sending buffers depends on the shifting communication windows $W$.
The window $W\left( N_{leap},NB_i \right)$ ($i \in \{ 1, 2, ..., P_x\}$) is now stretched across two neighbor processes.
For the current process $i$, the sending buffer is divided into two blocks $S1_i$ and $S2_i$.
As a result, there are two send and receive operations for each process in this case to apply the leap-format scheme.
Note that there are only one send and one receive operation under the circumstance that the shifting window coincides with a single process.



\subsection{Optimizations}
\label{sec:Opt}

\subsubsection{Message Aggregation}
The dynamical core is mainly made up of the advection process ($L_F$) and the adaption process ($A_F$), which are generally computed separately.
During each iteration of the model simulation, $L_F$ is called for $3*M$ times in the dynamical core, while $A_F$ is called for $3$ times, where $M$ denotes the speed difference between the two processes.
Thus, the shifting leap-format communication calls can be assorted and aggregated according to the different processes and various leap-format patterns.


Based on the previous discovery, we identify the leap-format pattern for each variable engaged in point-to-point shifting communication.
The communication aggregation scheme of variables involved in the leap-format difference computation is shown in Table~\ref{tab:comm}.
According to the different processes, variables are divided into two portions for communication calls. 
When performing shifting leap communication, variables of the same difference term, such as $UT,VT,TT$ (tendencies of $U,V,T$) in the advection process, 
\begin{figure}[htbp]
  \centerline{\includegraphics[width=3.5in]{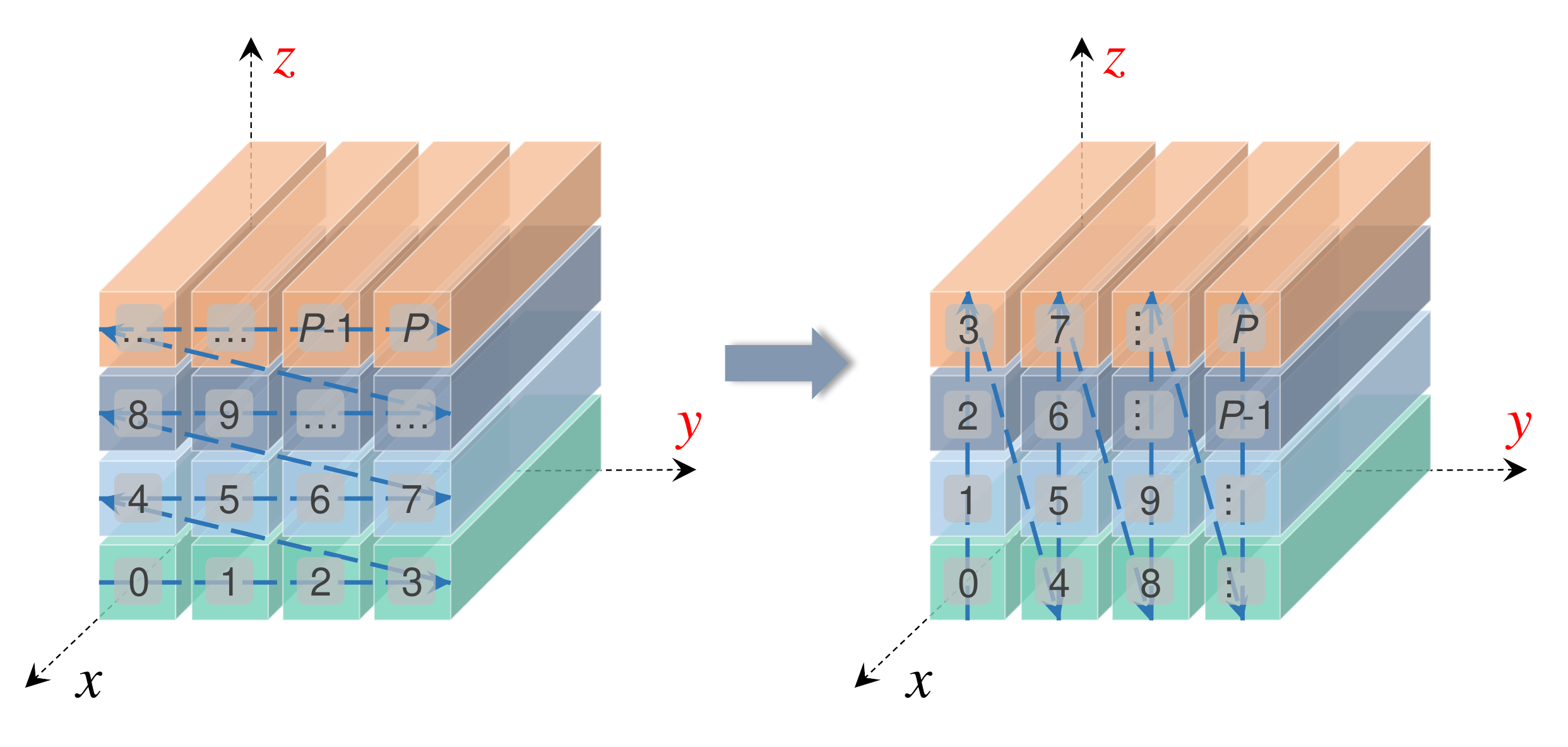}}
  \caption{Z-prior decomposition optimization strategy}
  \label{Z-prior}
\end{figure}
are collected in one send/receive buffer.
As a result, the message passing for shifting communication can achieve improved MPI bandwidth utilization.

Nonetheless, the huge quantity of communication volume and the use of computation/communication overlap should also be taken into full account before the communication aggregation of the same patterns.

\subsubsection{Decomposition Optimization Strategy}
As a typical communication-bound application, the cost of collective communication in IAP-AGCM dominates the overall execution time when performing large-scale simulations.

Most of the costly collective communication appears in the $Z$ sub-communicator for vertical computations.
For the dynamical core of IAP-AGCM, the MPI domain decomposition is on the basis of the three-dimensional Cartesian coordinates of the latitude-longitude mesh.
In the original version of the model, the $Y$ sub-communicator is preferentially deccomposed from the global communicator, then the $Z$ follow. To put it another way,
The location of a particular mesh point in different subdomains can be marked as $(L_x, L_y, L_z, L_a)$, 
where $L_x$, $L_y$, $L_z$ represents the process id 
\begin{table}[hbtp]
    \caption{Communication aggregation of leap-format difference variables}
    \begin{center}
    \begin{tabular}{m{1cm} m{2cm}<{\centering} m{2.7cm}<{\centering} m{1.2cm}<{\centering}}
    \toprule 
    \textbf{Process} & \textbf{Original difference terms}& \textbf{Leap-format difference terms} & \textbf{Variables} \\
    \midrule 
    \multirow{3}{*}[-14pt]{\textbf{Adaption}}&   $(x+1,\;x)$&   $(x+N_{leap},$ $x-N_{leap}+1)$&   $PXW, UT$\\ 
                                &   $(x,\;x-1)$&   $(x+N_{leap}-1,$ $x-N_{leap})$&   $PT, Pstar1$ $Pstar2, TT$ $deltap, GHI$\\ 
                                &   $(x+1,\;x-1)$& $(x+2*N_{leap}-1,$ $x-2*N_{leap}+1)$&   $Pstar2$   \\ 
    \midrule 
    \multirow{3}{*}[-10pt]{\textbf{Advection}}&  $(x+1,\;x)$&   $(x+N_{leap},$ $x-N_{leap}+1)$&   $Ustar$\\ 
                                        &   $(x,\;x-1)$&   $(x+N_{leap}-1,$ $x-N_{leap})$&   $Ustar$\\ 
                                        &   $(x+1,\;x-1)$& $(x+2*N_{leap}-1,$ $x-2*N_{leap}+1)$&   $UT,VT$ $TT$   \\ 
    \bottomrule
    \end{tabular}
    \label{tab:comm}
    \end{center}
\end{table}
\begin{figure}[htbp]
  \centerline{\includegraphics[width=3.5in]{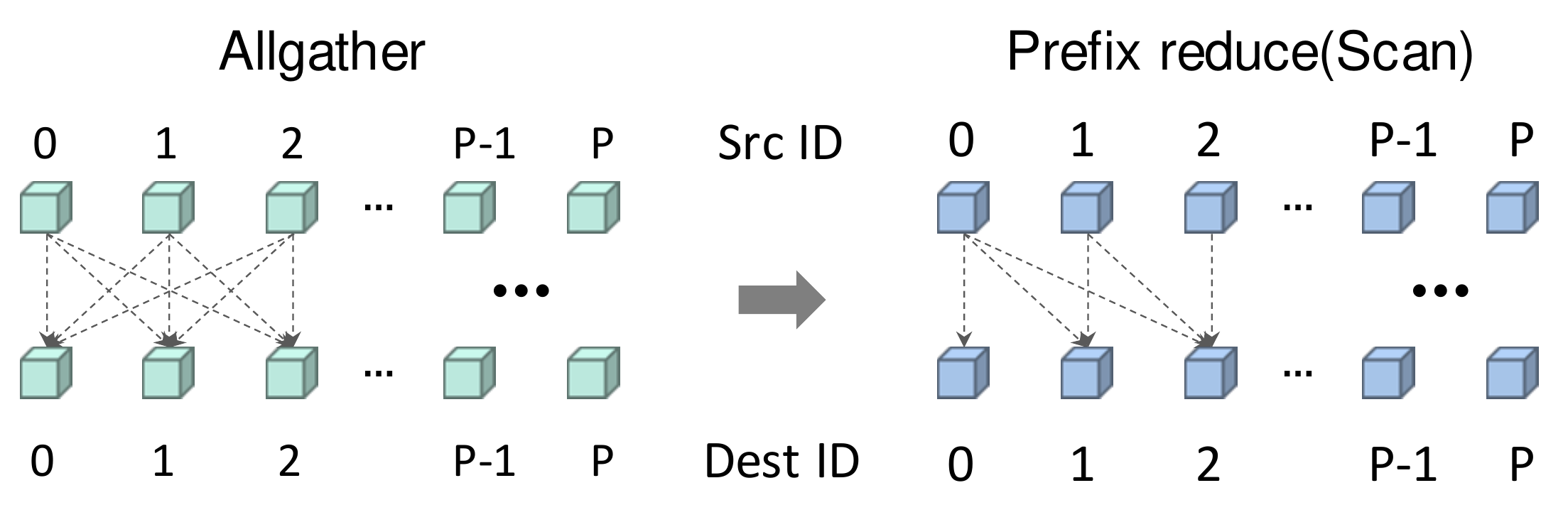}}
  \caption{Comparison of Communication operations}
  \label{scan}
\end{figure}
in the corresponding sub-communicators $X$, $Y$, $Z$, and $L_a$ is the rank in the global communicator. For simplicity, we omit the $X$ direction in the next discussion.
the location sequences $(L_y, L_z, L_a)$ for successive mesh points in different processes are 
$(0, 0, 0)$, $(1, 0, 1)$, $(2, 0, 2)$, ..., 
$(P_y-1, 0, P_y-1)$, $(P_y-1, 1, P_y)$,...,
$(P_y-1, P_z-1, (P_y \times P_z)-1)$. 
Take a modern multi-core processor/node in most supercomputers for example, 
this straightforward decomposition leads to the situation where there are no adjacent two mesh points along the $Z$ direction in one node 
when the parallelism along the $Y$ direction is greater than the number of cores per node. 
As a consequence, the collective communications along the $Z$ direction are all performed through the inter-node network, 
which brings in far more overhead than intra-node communication.

To mitigate the effect of the costly collective communication, we adopt a Z-prior decomposition optimization strategy to utilize the intra-node shared memory for MPI operations.
As shown in Fig.~\ref{Z-prior}, the prior decomposed sub-communicator is changed from $Y$-$Z$-$X$ to $Z$-$Y$-$X$.
The relevant collective communication operations along the $Z$ direction are then performed in physically affinitive processes.
Generally, the number of vertical layers in IAP-AGCM does not exceed the number of cores per node, which can assure that all communication operations along the $Z$ sub-communicator are intra-node.
Thus, the communication overheads for large-scale simulations can be greatly reduced.

\subsubsection{Refactoring of Computation and Communication}
In common with many other global atmosphere models, vertical velocity in IAP-AGCM is diagnosed from the continuity equation under hydrostatic equilibrium, which requires vertical integration from the surface ($\sigma$=1) to $\sigma$ level, thus causing many-to-many communications along vertical levels.

For the purpose of reducing the call frequency of MPI communication, 
the conventional parallel algorithm performs only one call of \emph{Allgather} in every iteration to acquire all needed data before computations.Take
the computation of the $\sigma$-surface vertical velocity ($WS$) for example, as shown in line 4 and line 8 of Algorithm~\ref{alg-1}, 
the sum and prefix sum operations of $\varepsilon$ and $\sigma^{L}(k)$ along the vertical dimension are executed after the call of \emph{MPI\_Allgather} in line 2.
The computations are performed covering all vertical layers in each process. Line 6 updates the intermediate variable $\delta^{L}(k)$ from $1$ to $L$.
And the sigma-surface vertical velocity $\sigma(k)$ is finally derived in line 10.
One benefit of this parallel algorithm is that the direct implementation is efficient enough for simulations on a smaller scale.
\begin{algorithm}[t]
  \SetKwInOut{Input}{Input}\SetKwInOut{Output}{Output}
  \newcommand\mycommfont[1]{\footnotesize\ttfamily\textcolor{blue}{#1}}
  \SetCommentSty{mycommfont}
  \SetKwBlock{DoParallel}{do in parallel}{end}
  
  \tcp*[h]{Divergence of (PT*VT),Coefficients,Number of vertical layers}\\
  \Input{$\delta$, $coef$, $L$ }  
  \Output{$\sigma$\tcp*[h]{Sigma-surface vertical velocity}}

  \BlankLine
  
  \DoParallel{
    \tcp*[h]{Allgather along $Z$}\\
    MPI\_Allgather($\delta$ , ..., $\delta^{L}$, ...)\; 
    
    \For{$k \leftarrow 1$ \KwTo $L$}{
      $\varepsilon \leftarrow \varepsilon + \delta^{L}(k) * coef(k)$;  \tcp*[h]{Vertical sum}\\
    }  
    \For{$k \leftarrow 1$ \KwTo $L-1$}{
      $\delta^{L}(k) \leftarrow \delta^{L}(k) + \varepsilon$; 
    }
    
    \For{$k \leftarrow 2$ \KwTo $L$}{
      $\sigma^{L}(k) \leftarrow \sigma^{L}(k-1) + \delta^{L}(k)$; \tcp*[h]{Prefix sum}
    }
    \For{$k \leftarrow beg$ \KwTo $end+1$}{
      $\sigma(k) \leftarrow \sigma^{L}(k)$; 
    }
  }
  \caption{Original implementation for computation of $\sigma$-surface vertical velocity.}
  \label{alg-1}
\end{algorithm}
\begin{figure}[htbp]
  \centerline{\includegraphics[width=3.5in]{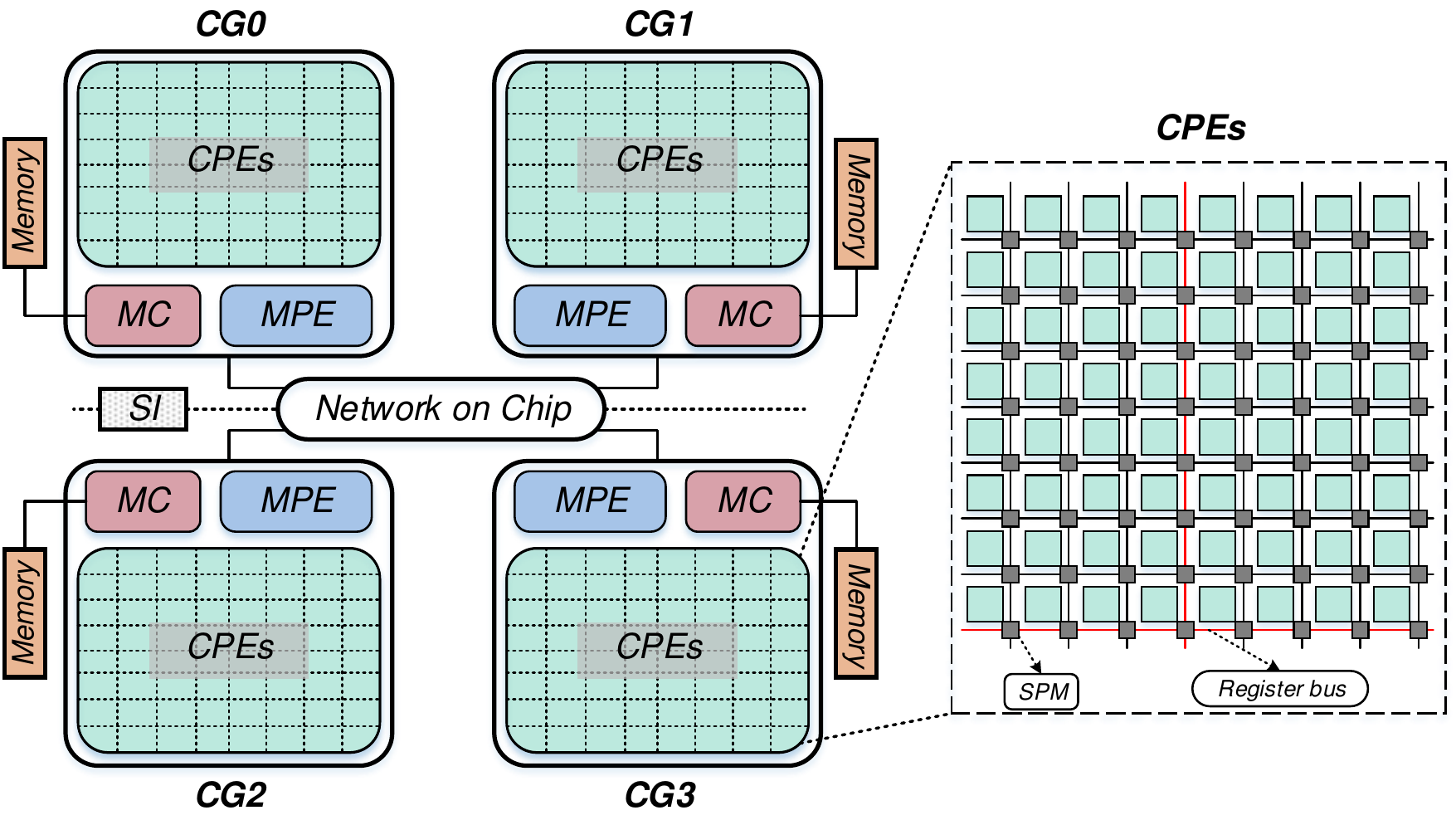}}
  \caption{The general architecture of SW26010 processor}
  \label{SW}
\end{figure}
However, with the increasing number of processes in large-scale simulations, the overhead of \emph{Allgather} communication
gradually dominates the overall execution time.
Also, the redundant computations in each process make Algorithm~\ref{alg-1} not the optimal choice.

To alleviate the disadvantages of the original design, we seek to refactor the relevant computations and communications as exhibited in Algorithm~\ref{alg-2}.Firsly
in Lines 2-4, the prefix sum ($\sigma$) and the vertical sum ($\varepsilon$) are calculated locally from $beg$ to $end$.
Line 5 and 6 perform the \emph{MPI\_Allreduce} (sum) and \emph{MPI\_Exscan} (prefix sum) communications along vertical levels, respectively.
The intermediate variables $\varepsilon^{sum}$ and $\varepsilon^{scan}$ are then added to $\sigma(k)$ from locally to obtain the final result in Line 8.
Overall, the sum and prefix sum operations are reconstructed and executed locally.
Whereas one extra call of collective communication is introduced in the new design, the whole communication volume decreases significantly, as shown in Fig.~\ref{scan}.
Moreover, no redundant computations except the ghost zones are executed among different processes along $Z$.
In addition, it will be more feasible to utilize non-blocking collective communications for overlapping some overhead when adopting the refactoring algorithm.
In summary, the new design achieves a tradeoff between the volume, the call frequency of MPI communication, and the redundant computations.


\begin{algorithm}[t]
  \SetKwInOut{Input}{Input}\SetKwInOut{Output}{Output}
  \newcommand\mycommfont[1]{\footnotesize\ttfamily\textcolor{blue}{#1}}
  \SetCommentSty{mycommfont}
  \SetKwBlock{DoParallel}{do in parallel}{end}

  \tcp*[h]{Divergence of (PT*VT),Coefficients}\\
  \Input{$\delta$, $coef$}  
  \Output{$\sigma$  \tcp*[h]{Sigma-surface vertical velocity}}

  \BlankLine
  
  \DoParallel{
    \tcp*[h]{Local sum/prefix sum along $Z$}\\

    \For{$k \leftarrow beg$ \KwTo $end$}{
      $\sigma(k+1) \leftarrow \sigma(k) + \delta(k) \ast coef(k)$\; 
      $\varepsilon \leftarrow \varepsilon + \delta(k) \ast coef(k)$\; 
    } 
    \tcp*[h]{Allreduce(Sum) along $Z$}\\ 
    MPI\_Allreduce($\varepsilon$, ..., $\varepsilon^{sum}$, ..., $sum$, ...)\;
    \tcp*[h]{Scan(Prefix sum) along $Z$}\\ 
    MPI\_Exscan($\varepsilon$, ..., $\varepsilon^{scan}$, ..., $sum$, ...)\;

    \For{$k \leftarrow beg$ \KwTo $end+1$}{
      $\sigma(k) \leftarrow \sigma(k) + \varepsilon^{sum} + \varepsilon^{scan}$\; 
    }
  }
  \caption{Refactoring implementation for computation of $\sigma$-surface vertical velocity.}
  \label{alg-2}
\end{algorithm}
 
\subsection{Heterogeneous Acceleration}
\label{sec:SW}
\begin{figure}[htbp]
  \centerline{\includegraphics[width=3in]{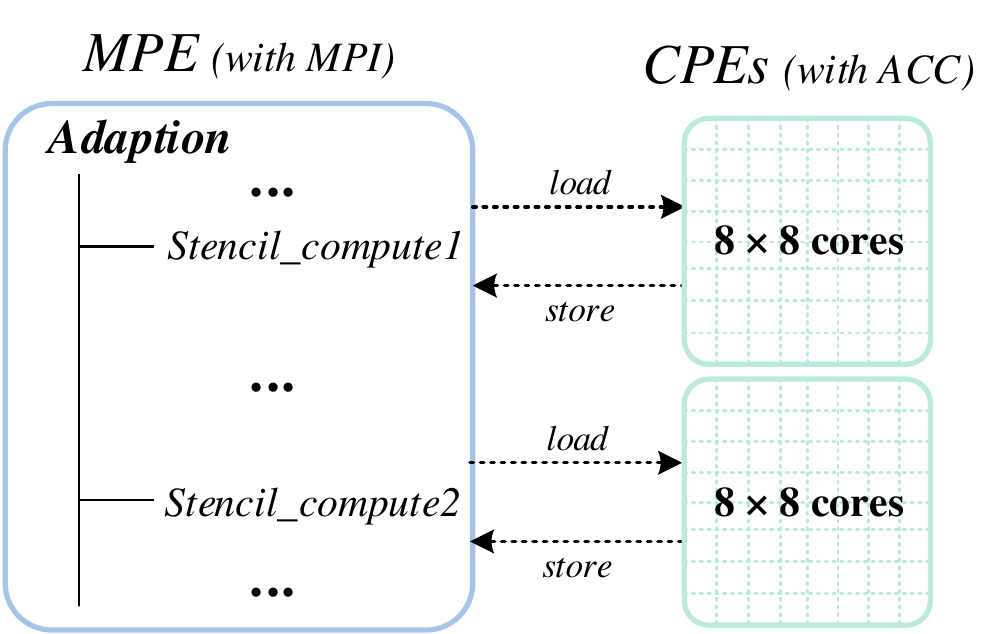}}
  \caption{The two-level heterogeneous parallelization.}
  \label{ACC}
\end{figure}

To analyze the suitability for heterogeneous HPC platforms, we deploy the model to a heterogeneous supercomputer, the Sunway TaihuLight.
As a milestone for Chinese homegrown supercomputers, the Sunway TaihuLight system claims 4th place in the latest TOP500 list.
The whole system scales to 10,649,600 cores (40,960 supernodes), with a sustained Linpack performance of 93 PFLOPS and a peak performance of over 125 Pflops.
TaihuLight is built using the heterogeneous many-core processor, SW26010, whose architecture is shown in Fig~\ref{SW}. 
The SW26010 processor makes up of 4 core groups (CGs), each of which contains one management processing element (MPE) and 64 computing processing elements (CPEs).
The main memory of one CG is interconnected with the MPE and CPEs by the memory controller (MC).
And each CPE possesses a 16 KB L1 instruction cache and a 64 KB Scratch Pad Memory (SPM).

Based on the characteristic of IAP-AGCM and TaihuLight, we employ a two-level MPI\,+\,OpenACC heterogeneous scheme to utilize the acceleration resources.
Take the computation in the adaption process for example, firstly, the MPE copies the relevant data of stencil computations on the MPI process to the CPEs via DMA.
After decompression conducted by the messenger CPEs, the needed data is transferred to other CPEs for stencil computations through register communication.
The computation result is then transferred back to the MPI process.
Fig.~\ref{ACC} shows the whole procedure.
Considering that the computation hotspots in IAP-AGCM are scattered during the iteration, we individually handle the stencil loops to leverage the CPEs.
To maximize the performance improvement for each stencil loop, the DMA transfer tile should be taken into consideration for the utilization of the local direct memory.

Algorithm~\ref{alg-3} exhibits the acceleration for the computation of a 3D variable $DIV$.
Lines 1-3 and Line 8 explicitly specify the OpenACC directive clauses.
In detail, \emph{parallel loop} and \emph{end parallel loop} represent the beginning and end of the offloading acceleration region.
\emph{collapse} is the number of fused parallelizing loops, which is 2 in the case of Algorithm~\ref{alg-3}.
\emph{tile} indicates the blocking size for DMA transfer. For different stencil computations, the \emph{tile} size should be accordingly adjusted.
\emph{copyin, copyout} denote the input (load) and output (store) of the data transfer.
And \emph{annotate} annotates the properties and dimensions, etc. of related variables for the correct processing of the ACC compiler.

By explicitly specifying the parameters in the directive clauses, the ${\rm Stencil}(\alpha_{i,j}, U_{i,k,j}, V_{i,k,j})$ computation kernel is loaded to the CPEs for acceleration, which is the basic procedure for a large number of stencil loops' acceleration.
Throughout the adaption and the advection process in IAP-AGCM, more than 30 stencil loops are heterogeneously parallelized, and an individually 3x$\thicksim$20x performance speedup is derived from the MPI\,+\,OpenACC hybrid parallelization.

\begin{algorithm}[t]
  \SetKwInOut{Input}{Input}\SetKwInOut{Output}{Output}
  \newcommand\mycommfont[1]{\footnotesize\ttfamily\textcolor{blue}{#1}}
  \SetCommentSty{mycommfont}
  \SetKwBlock{DoParallel}{do in parallel}{end}

  \tcp*[h]{Tendencies and auxiliary variables}\\
  \Input{$\alpha, U, V, ...$}  
  \Output{$DIV$ } 

  \BlankLine
  
    \tcp*[h]{OpenACC directive clauses}\\
    \emph{{!\$ACC parallel loop collapse(2) tile(k:4) local(i,k,j) {\rm\&}}}\\
    \emph{{!\$ACC copyin($\alpha, U, V$) copyout($DIV$) {\rm\&}}}\\
    \emph{{!\$ACC annotate(readonly(...); dimension(...))}}\\
    \For{$j \leftarrow beg\_y$ \KwTo $end\_y$}{
      \For{$k \leftarrow beg\_z$ \KwTo $end\_z$}{
        \For{$i \leftarrow beg\_x$ \KwTo $end\_x$}{
          \tcp*[h]{Compute Divergencies \& their sums}\\
          $DIV_{i,k,j} \leftarrow  {\rm Stencil}(\alpha_{i,j}, U_{i,k,j}, V_{i,k,j})$\;
        } 
      }
    }
    \tcp*[h]{End of directive clauses}\\ 
    \emph{{!\$ACC end parallel loop}}\\


  \caption{CPEs acceleration for the computation of $DIV$.}
  \label{alg-3}
\end{algorithm}

\section[]{Evaluation}
\label{sec:perf}
In this section, we present the performance results of our new 3D domain decomposition method and the leap-format scheme. 
The correctness of the new implementation is verified by several waveform tests.

\subsection{Platform and Setup}
\begin{table*}[htbp]
  \caption{Processes configurations for different resolutions of 2D and 3D decomposition}
  \begin{center}
  \begin{tabular}{m{2.5cm}<{\centering} m{2.5cm}<{\centering} m{2.5cm}<{\centering} m{2.2cm}<{\centering} m{2.5cm}<{\centering} m{2.4cm}<{\centering} }
\toprule 
   {\textbf{horizontal}} & \textbf{Mesh points} & \textbf{2D decomposition} & {\textbf{2D max}} & {\textbf{3D decomposition}} &  {\textbf{3D/Leap max }} \\
  \textbf{resolution} & \textbf{($N_x \times N_y \times N_z$)} & \textbf{($P_y \times P_z$)} & \textbf{parallelism} & \textbf{($P_x \times P_y \times P_z$)} & \textbf{parallism}\\
\midrule
  $1.4\degree\times1.4\degree$   &   $256\times  128\times 30$ & $64 \times  16$ & 1,024 & $16 \times 64 \times  16$ & 4,096  \\
  $0.5\degree\times0.5\degree$   &   $720\times  361\times 30$ & $180 \times 16$ & 2,880 & $16 \times 180 \times 16$ & 32,768  \\
  $0.25\degree\times0.25\degree$ &   $1152\times 768\times 30$ & $384 \times 16$ & 6,144 & $32 \times 384 \times 16$ & 196,608 \\
\bottomrule   
  \end{tabular}
  \label{tab:processes}
  \end{center}
\end{table*}

Our simulation tests are performed on the Supercomputer CAS-Xiandao1 and Sunway TaihuLight. Each computational node of CAS-Xiandao1 has one Hygon Dhyana 
\begin{figure}[htbp]
  \centering
    \subfigure[R-H waveform for the AGCM dynamical core.]{
      \label{R-Ha} 
      \includegraphics[width=2in]{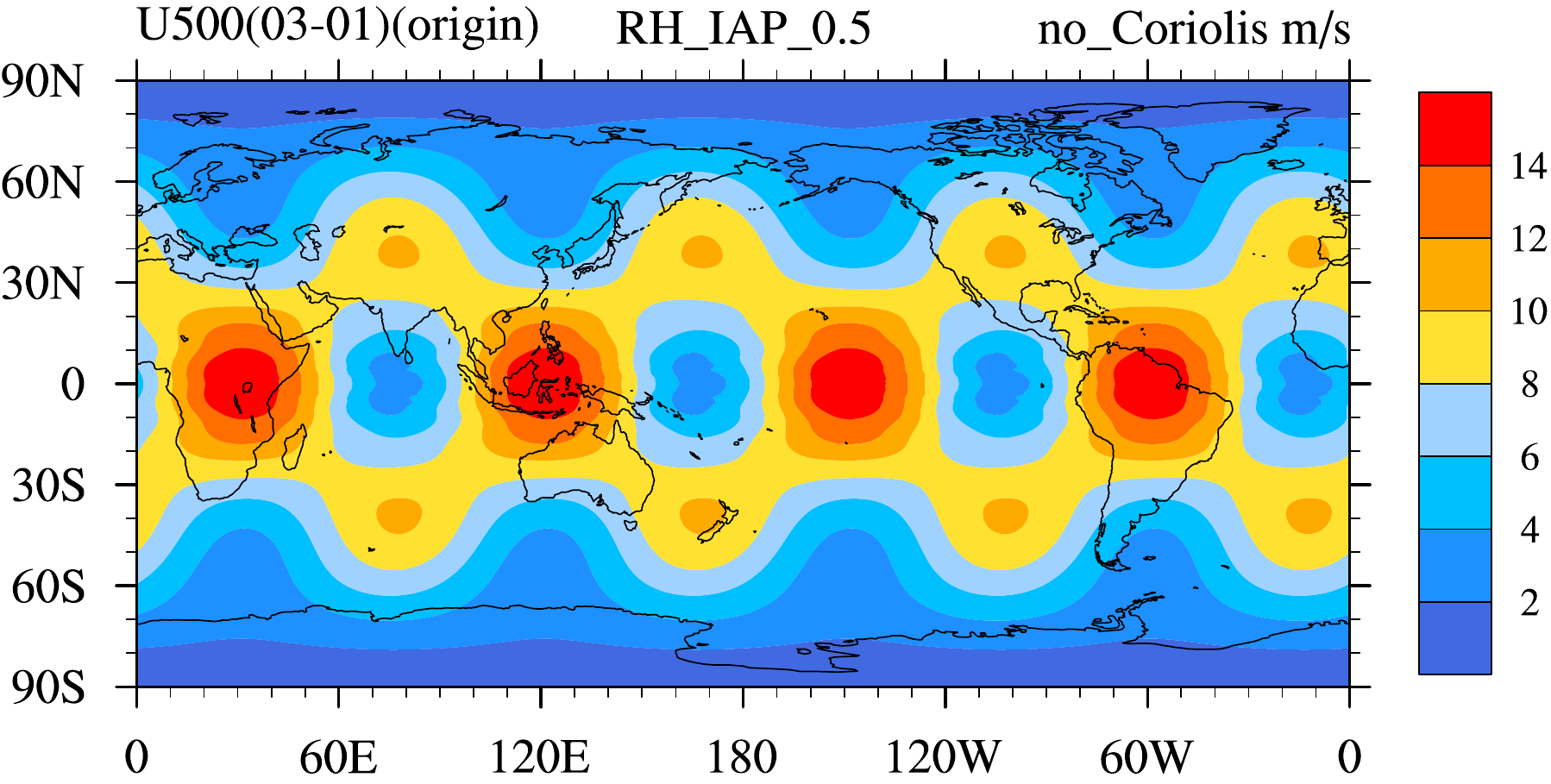}} \\
    \subfigure[Surface temperature for AGCM-3DLF.]{
      \label{R-Hb} 
      \includegraphics[width=1.6in]{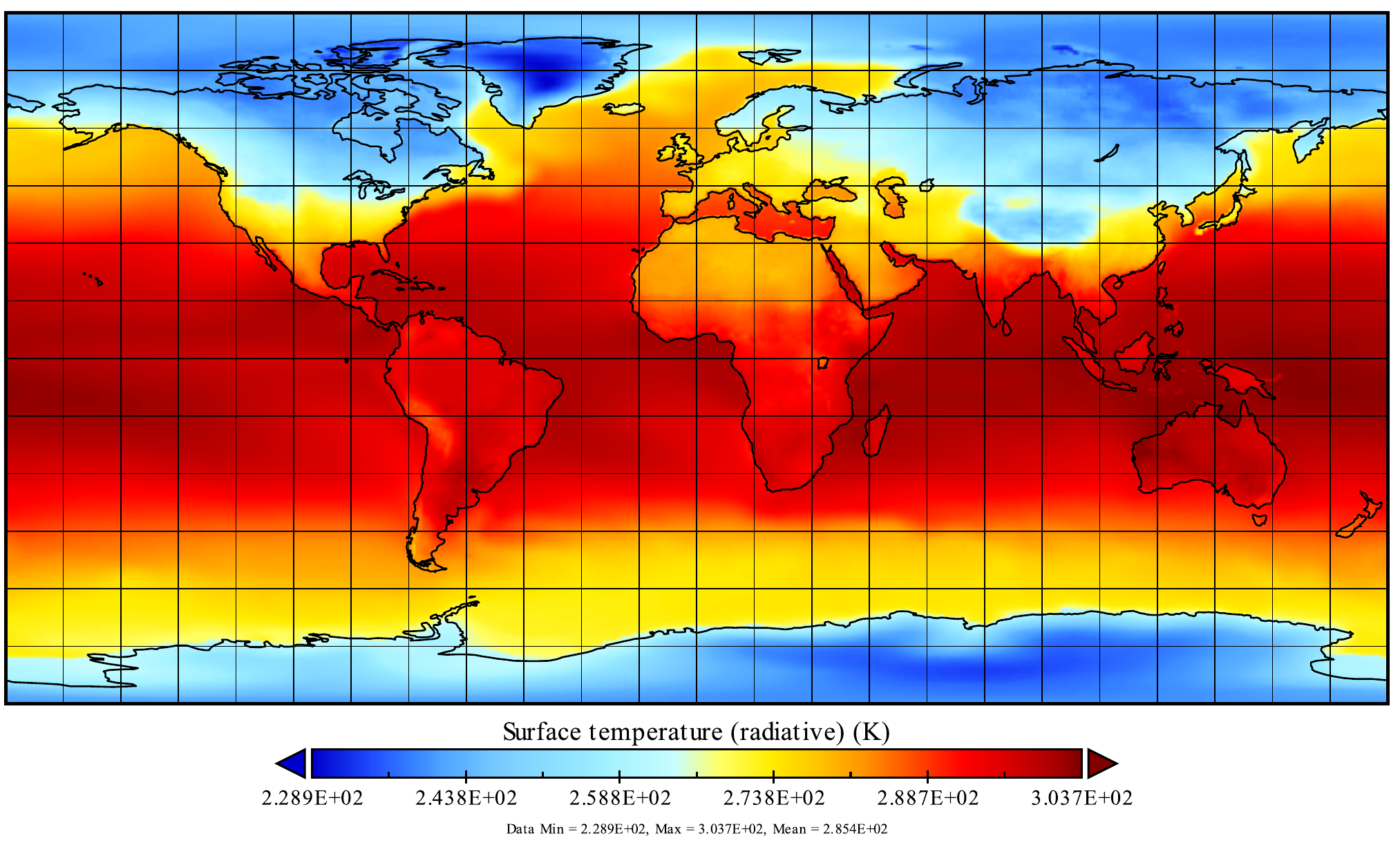}}
    \subfigure[Total (convective and large-scale) precipitation rate for AGCM-3DLF.]{
        \label{R-Hc} 
        \includegraphics[width=1.6in]{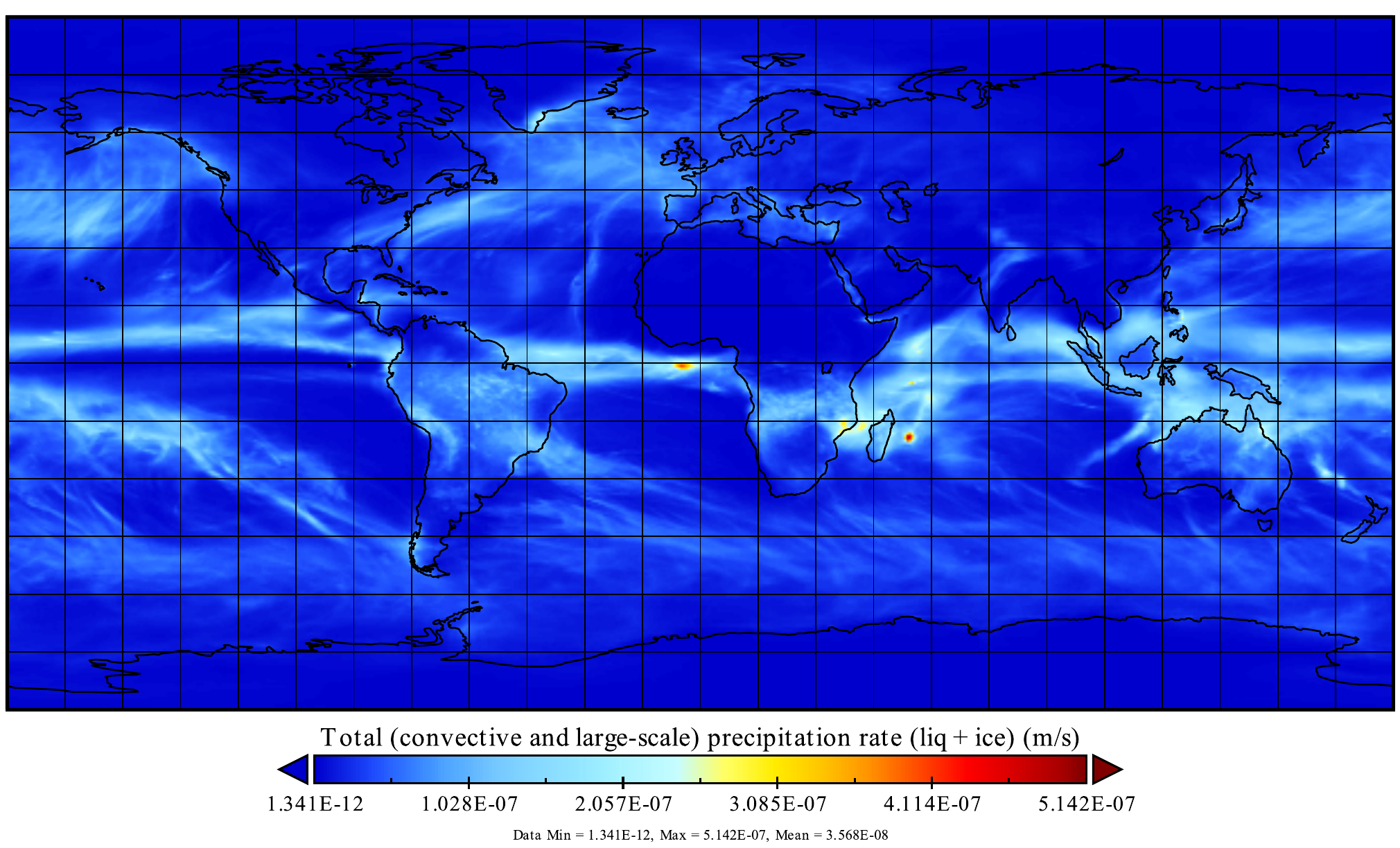}}
    \caption{R-H 4 waves test for zonal wind $U$ and F\_2000 tests for surface temperature $T$ and total precipitation rate. The distributions of the R-H wave are derived from the output data of 2 simulated months. The test aims to examine the impact of the spherical baroclinic dynamical core without moist physics.}
  \label{R-H}
\end{figure}
\begin{figure}[htbp]
  \centerline{\includegraphics[width=3.2in]{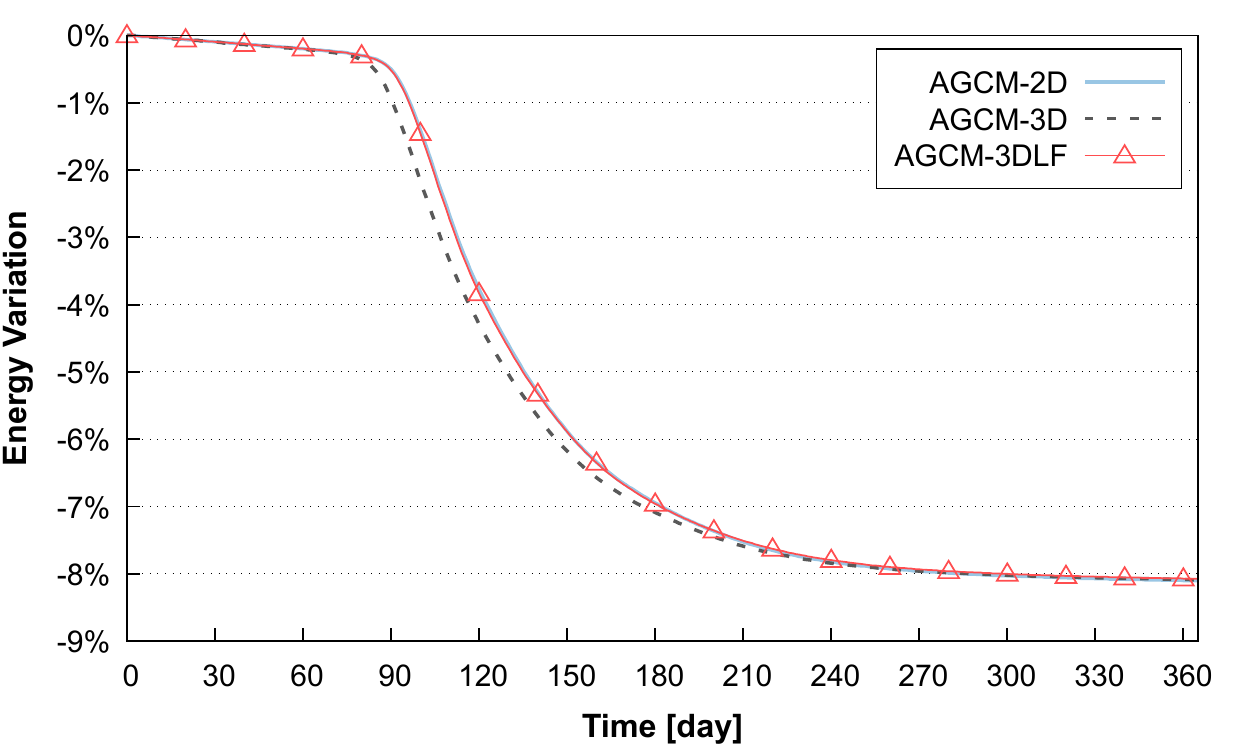}}
  \caption{The total global mean energy conservation of three implementations. The total effective energy makes up of the kinetic energy, the available potential energy, and the available surface potential energy.}
  \label{Energy_total}
\end{figure}
processor (total 32 cores) with 128 GB memory. The computing nodes are interconnected with a 200GB/s InfiniBand network. The system software includes the backend compiler Intel 17.0, and a customized communication library HPC-X MPI.


To analyze the performance of the model thoroughly, we choose different resolution options from moderate $1.4\degree\times1.4\degree$ to finer $0.25\degree\times0.25\degree$ supported in the current version of IAP-AGCM. The vertical layer ($Z$) is set to $30L$.
For different resolution configurations, the number of mesh points varied. The detailed setup of mesh points along with the parallelism is listed in Table~\ref{tab:processes}.

As can be seen, to achieve the highest simulation performance, the group of processes is distributed in three dimensions and scales to the largest and feasible number accordingly in AGCM-3D/AGCM-3DLF. 
\begin{figure}[htbp]
  \centerline{\includegraphics[width=3.2in]{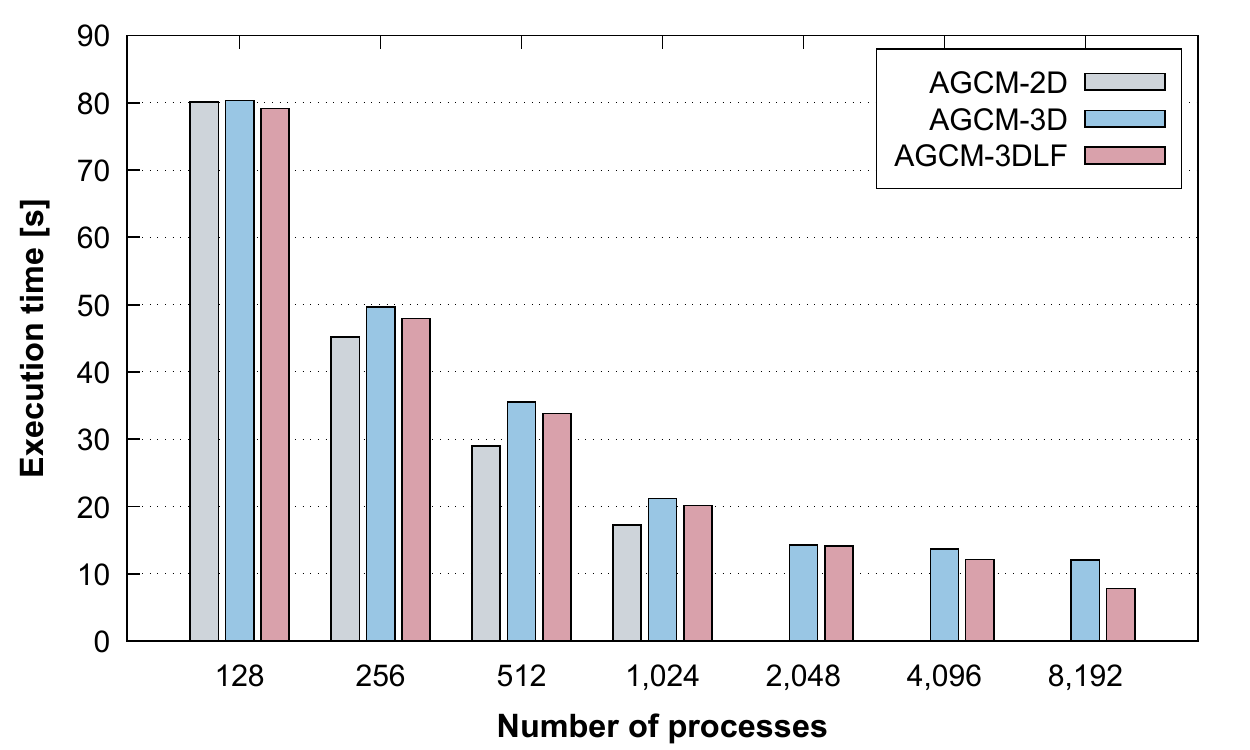}}
  \caption{Overall execution time comparison for different implementation at the horizontal resolution of $1.4\degree\times1.4\degree$.}
  \label{Perf_runtime}
\end{figure}
In our experiments, the maximum number of processes utilized for both the original and leap-format difference dynamical cores is 196,608.
A series of \emph{F\_2000} coupled model (IAP-AGCM coupling with land model) tests are conducted as the strong scaling tests for the whole model.


\subsection{Validation of AGCM-3DLF}

The Rossby-Haurwitz (R-H) waves test is a widely used test method as the exact solutions of the spherical barotropic vorticity equation~\cite{phillips1959numerical,xuehong1990dynamical} for an ideal and unforced fluid in the meteorological fields.
To validate the simulation results of the AGCM-3DLF, the R-H test is adopted for the dynamical core along with the F\_2000 coupled AGCM model simulations.

We conduct a set of validation tests for our new implementation.
The waveform of zonal wind $U$(m/s) and the coupled simulation results are presented in Fig.~\ref{R-H}.
As shown in Fig.~\ref{R-Ha}, The four R-H waveforms of the model maintain properly after 2 months' simulations. The distribution difference of $U$ is less than 0.1 m/s when comparing the waveforms with the naïve implementation. 
In Fig.~\ref{R-Hb} and Fig.~\ref{R-Hc}, the surface temperature and the total (convective and large-scale) precipitation rate for the coupled model are presented. 
In comparison with the observation data, both variables fit in expectation.

Another key indicator of the AGCM is energy conservation.
In Fig.~\ref{Energy_total}, we present the total global mean energy attenuation for the original AGCM-2D, the AGCM-3D, and the AGCM-3DLF, respectively.
In a latitude-longitude mesh-based model, the total effective energy should be conserved well in a short-term R-H test.
\begin{figure}[htbp]
    \centering
    \subfigure[Heatmap of the AGCM-3D implementation.]{
      \label{org_hm} 
      \begin{minipage}[b]{0.2\textwidth}
        \includegraphics[height=1.4in]{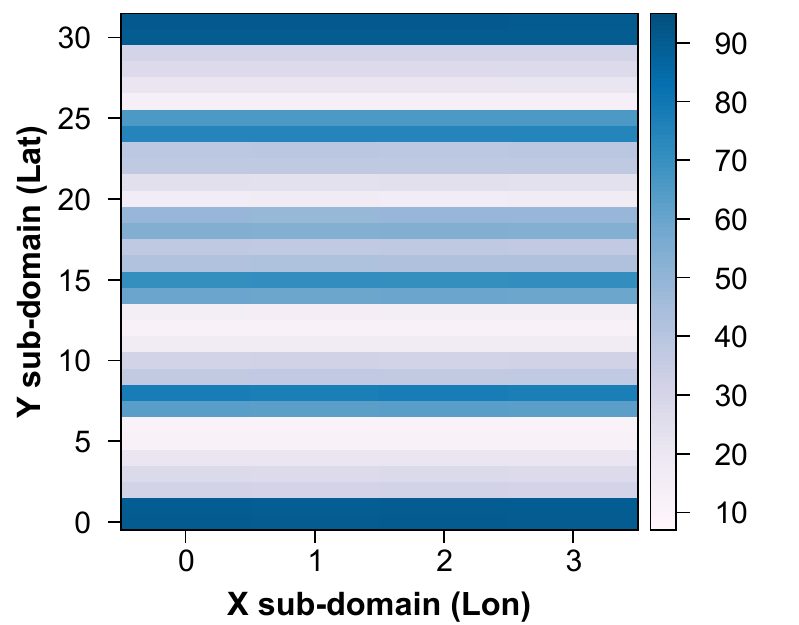}
      \end{minipage}
    }
    \hspace{3.5mm}
    \subfigure[Heatmap of the AGCM-3DLF with leap-format implementation.]{
      \label{leap_hm} 
      \includegraphics[height=1.4in]{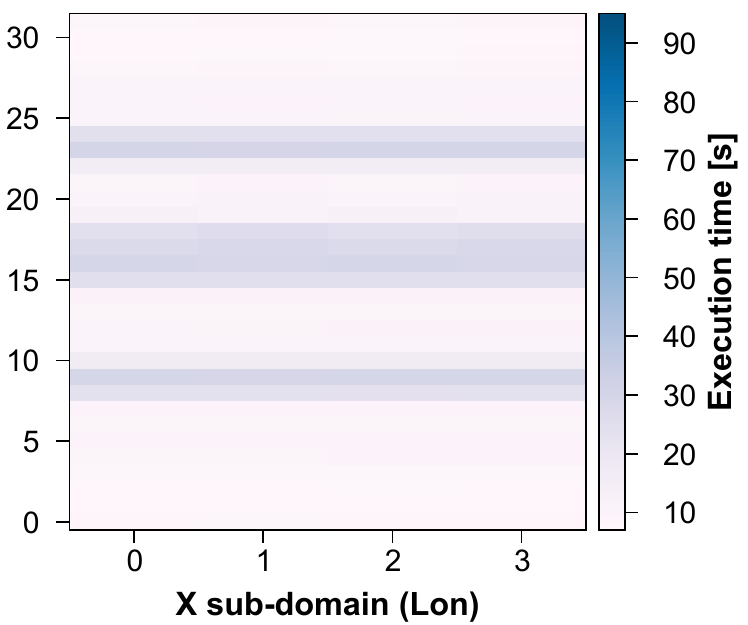}
    }
    \caption{The heatmaps of execution time of 128 processes for the AGCM-3D and the AGCM-3DLF.}
  \label{heatmap}
\end{figure}
As shown in the figure, the total energy attenuation of all three implementations is within the acceptable range in 90 days.
Note that AGCM-3DLF conserves energy better than the plain AGCM-3D from day 90 to day 180, which demonstrates the accuracy of the leap-format scheme.



\subsection{Effectiveness Measurements}

\subsubsection{3d Decomposition with Leap-format}

To demonstrate the viability of the AGCM-3DLF, we show the comparison of execution time for the original AGCM-2D, 
the AGCM-3D, and the AGCM-3DLF at the horizontal resolution of $1.4\degree\times1.4\degree$ in Fig.~\ref{Perf_runtime}. 
Though AGCM-2D slightly outperforms the 3D model in most cases from 128 processes to 1,024 processes, 
1,024 is the largest number of processes the 2D model can scale to, as seen in Fig.~\ref{Perf_runtime}.
In the range of 2,048 to 8,192 processes, the execution time of the AGCM-3D implementations is decreasing continually.
Thereinto, the performance of the leap-format model apparently offers large performance benefits, 
especially when the parallelism approaches the maximum.

The reason lies in the previous discussion in Section~\ref{sec:BG.Filt}.
Recall that serious load imbalance occurs in the filtering module, which hiders the scaling of the model.
Fig. \ref{heatmap} presents the heatmaps of the execution time for the
AGCM-3D and the AGCM-3DLF, which indicate the load balancing of the filtering module in the horizontal direction ($X$-$Y$).
The number of processes in the heatmaps is set to 128, and the decomposition is fixed at $X \times Y \times Z = 32 \times 4 \times 1$ for brevity.
Thus the number of grid points of the heatmaps is $32 \times 4$.
\begin{figure}[htbp]
  \centerline{\includegraphics[width=3.2in]{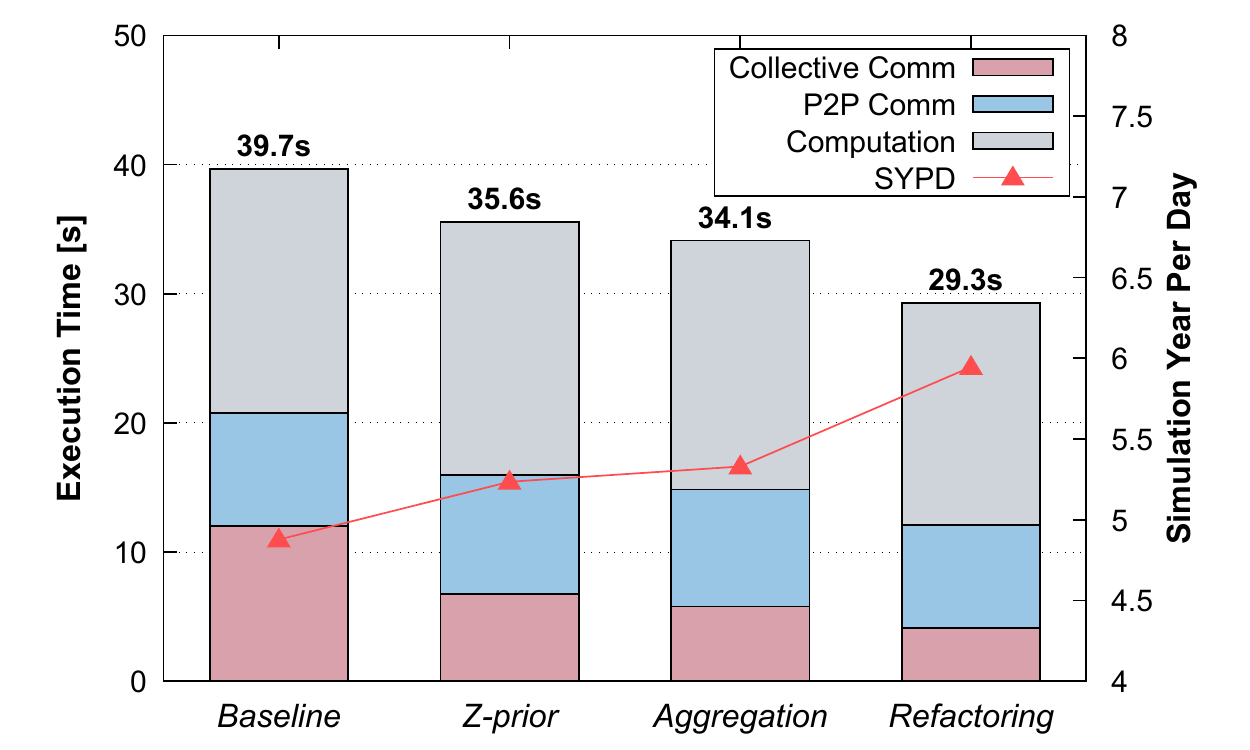}}
  \caption{Execution time comparison of the optimization techniques.}
  \label{opt}
\end{figure}
As can be seen, for both implementations, the overheads of filtering are basically balanced in different processes along the $X$ subdomain.
However, the execution time varies drastically along the $Y$ subdomain due to the adaptive Gaussian filtering for the AGCM-3D. 
The gap between the maximum and minimum overheads can be as high as 8.3x (90.5s compared to 10.8s). 
With regard to the AGCM-3DLF, the serious imbalance along the $Y$ subdomain is alleviated by the leap-format difference computation (29.8s compared to 7.6s).
A point of note is that the slightly higher execution time at the mid-latitudes in Fig.~\ref{leap_hm} is caused by the smoothing process, which also exists in Fig.~\ref{org_hm}.
To sum up, the leap-format brings in significant gains for the model's scaling.

\subsubsection{Optimizations}

As discussed in Section~\ref{sec:Opt}, several optimization techniques are utilized to achieve higher performance in the simulation.
Fig.~\ref{opt} exhibits the effectiveness of the optimizations for the parallelism of 6,144 processes at the resolution of $0.25\degree\times0.25\degree$.
The four bars in the histogram represent the execution time of the baseline, and different implementations added up with the optimization techniques successively.
With all three techniques, the execution times of communications are decreased. Especially, the collective communication overheads are significantly reduced. 
The overall speedup compared to the baseline is 1.36x. 
Note that for larger-scale simulation, the proportion of the computation tends to decline, which will emphasize the speedup of these optimizations.

\subsection{Scaling Tests}

\begin{figure}[tbp]
  \centering
    \subfigure[Scaling results for the resolution $1.4\degree$.]{
      \label{S_C_a} 
      \includegraphics[width=3.2in]{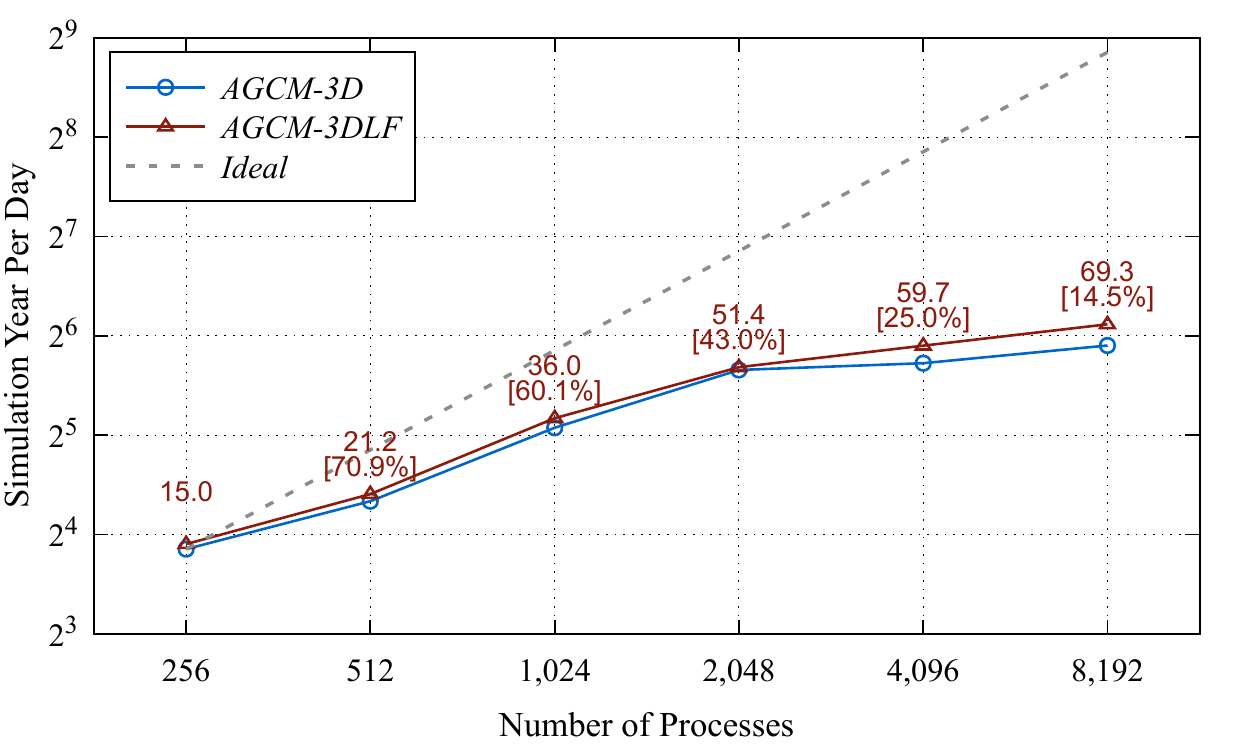}} \\
    \subfigure[Scaling results for the resolution $0.5\degree$.]{
      \label{S_C_b} 
      \includegraphics[width=3.2in]{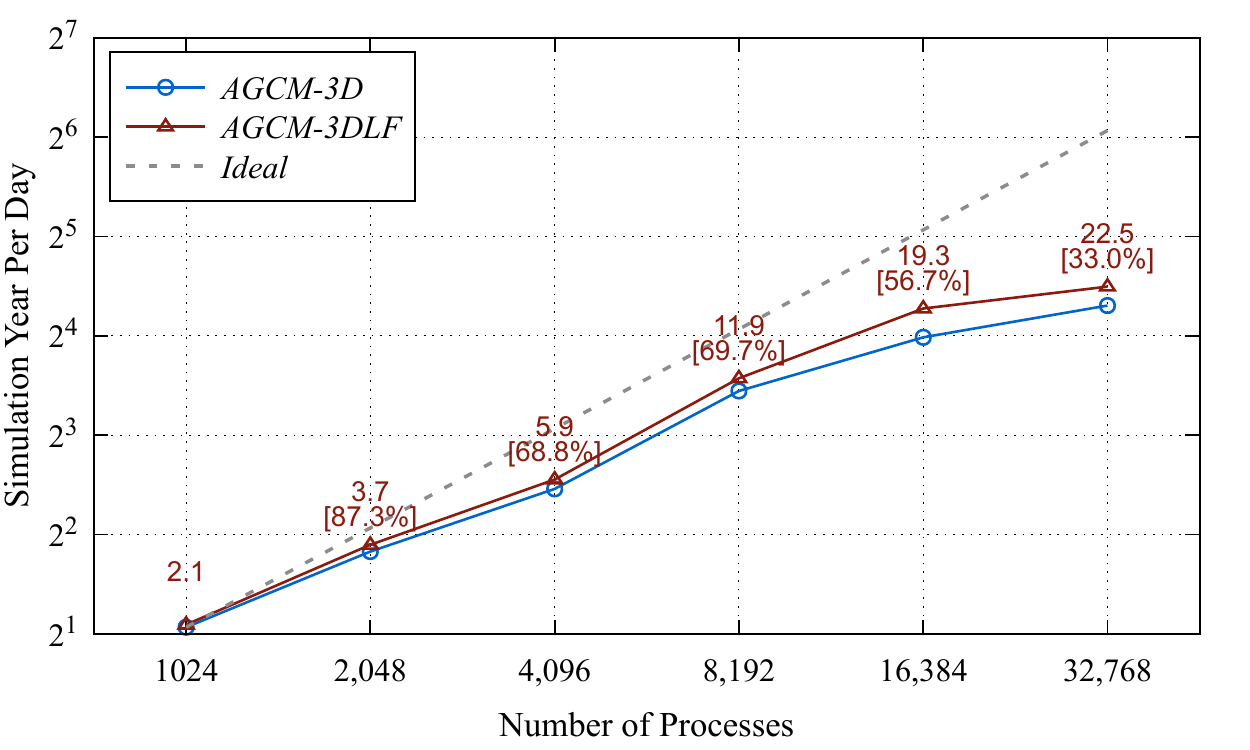}} \\
    \subfigure[Scaling results for the resolution $0.25\degree$.]{
        \label{S_C_c} 
        \includegraphics[width=3.2in]{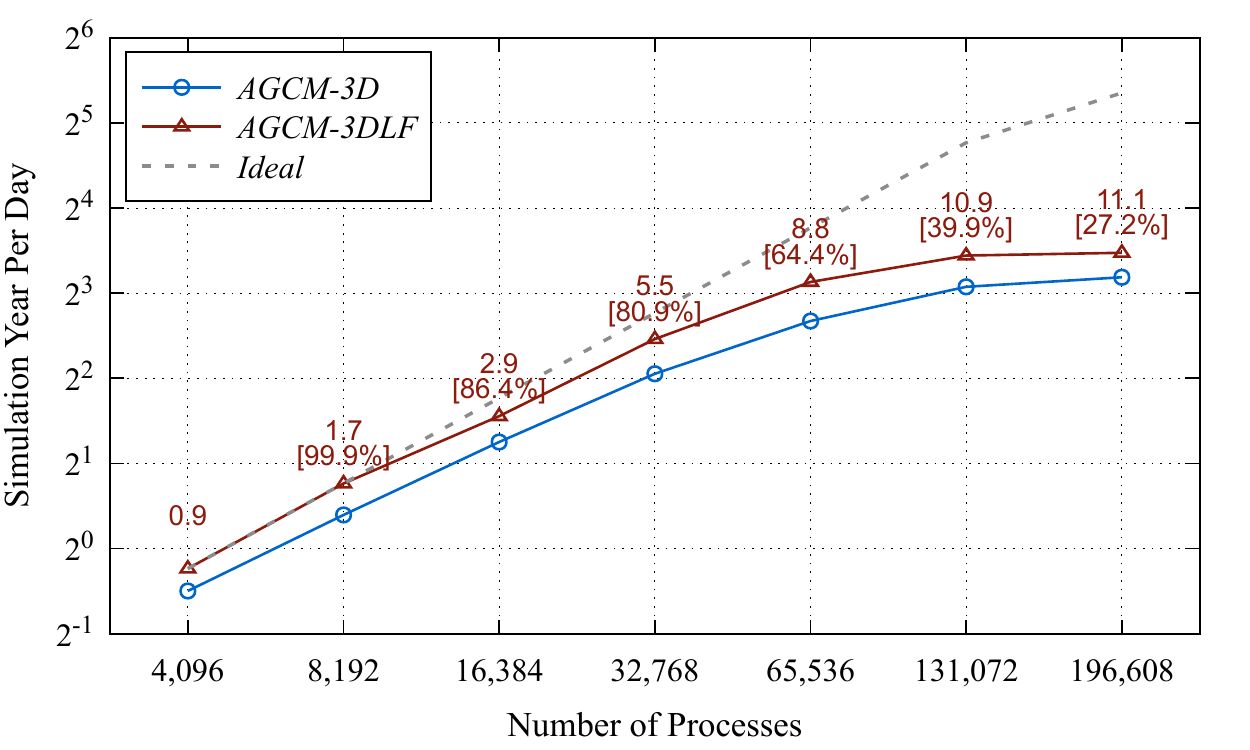}}
    \caption{Strong scaling results for different resolutions. The simulations are carried out with AGCM-3D and AGCM-3DLF. The computing throughput is measured in Simulation Year Per Day (SYPD). \emph{Ideal} represents the corresponding ideal performance.The percentages in square brackets [] indicate the parallel efficiency for different scales.}
  \label{strong_scaling}
\end{figure}

The strong scaling tests are performed at three different horizontal resolutions from $1.4\degree$ to $0.25\degree$ to investigate the overall scalability,
as shown in Fig. \ref{strong_scaling}. The detailed simulation configuration corresponds to Table \ref{tab:processes}. 
For the resolution $1.4\degree$, the model scales from 256 processes to 8,192 processes, obtaining a 4.6x speedup with the parallel efficiency of 14.5\%. 
For the resolution $0.5\degree$, the model scales from 1,024 processes to 32,768 processes, obtaining a 10.6x speedup with the parallel efficiency of 33.0\%. 
And for the resolution $0.25\degree$, the model scales from 4096 processes to 196,608 processes, i.e., the full system scale, and the simulation obtains a 13.1x speedup with the parallel efficiency of 27.2\%. The maximum computing throughputs measured in \emph{SYPD} are 69.3, 22.5, and 11.1 for the resolutions of $1.4\degree$, $0.5\degree$ , and $0.25\degree$, respectively.

The main reason for the inferior parallel efficiency at the resolution of $1.4\degree$ is that the number of mesh grids in one process is smaller than that of finer resolutions. And all three simulations with AGCM-3DLF achieve better performance than the plain AGCM-3D. 
For the resolutions of $1.4\degree$ and $0.5\degree$, AGCM-3DLF provides average speedups of 8\%$\thicksim$10\% over AGCM-3D.
\begin{figure}[htbp]
  \centerline{\includegraphics[width=3.2in]{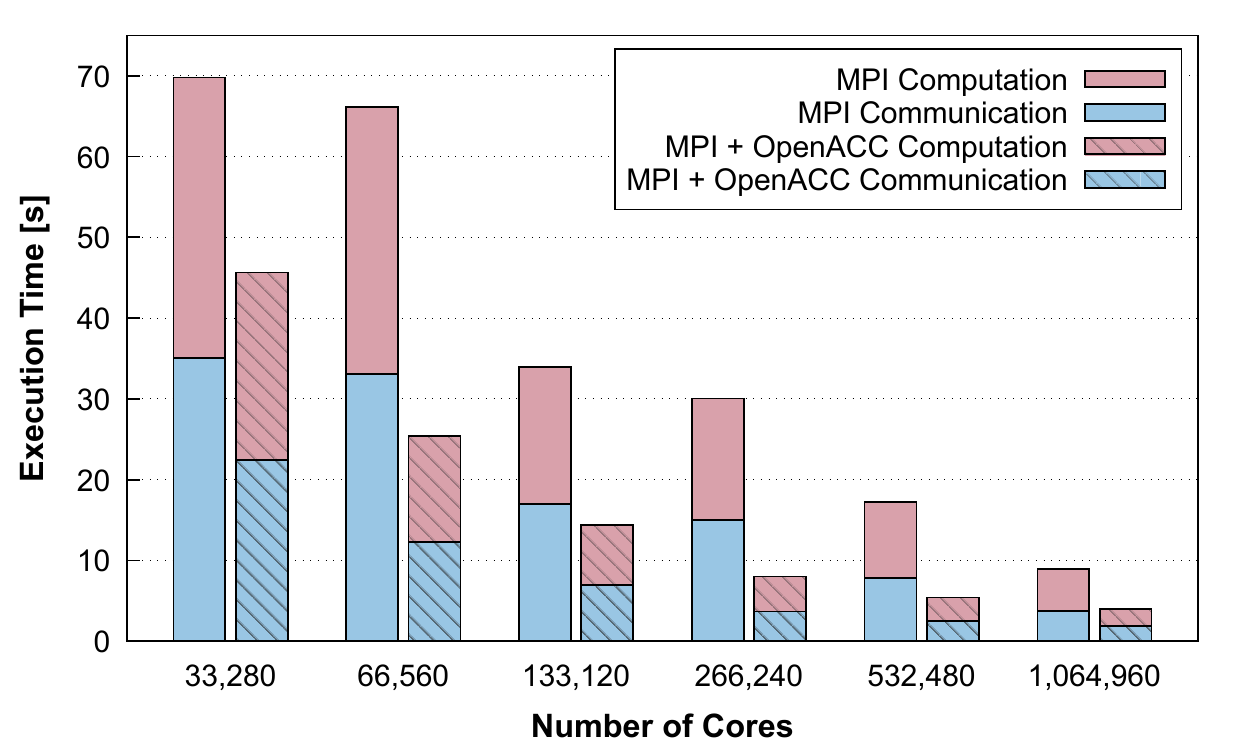}}
  \caption{Strong scaling results at the resolution of $0.25$ on Sunway TaihuLight.}
  \label{Perf_SW}
\end{figure}
In particular, the improvement for the $0.25\degree$ high-resolution model provides a 37.4\% boost of performance at large.
A point of note is that the performance results of AGCM-2D are not shown in the figures, 
because both the scalability and the sustained performance are not compatible, especially in high-resolution simulations.
In summary, AGCM-3DLF accomplishes essential performance promotion for all three resolutions on the Supercomputer CAS-Xiandao1.

\begin{figure}[htbp]
    \centerline{\includegraphics[width=3.2in]{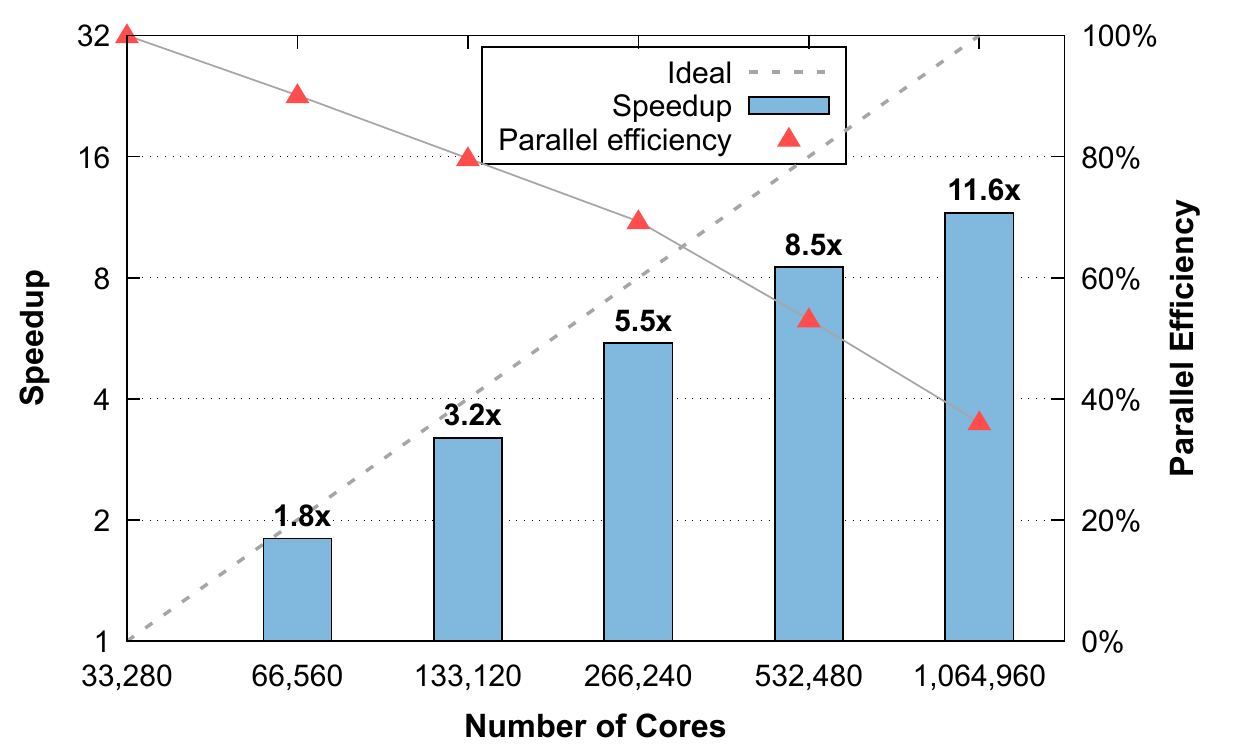}}
    \caption{Speedup and parallel efficiency of scaling results in Sunway TaihuLight. Ideal represents the corresponding ideal performance.}
    \label{speedup_SW}
\end{figure}

In addition, simulation results of AGCM-3DLF conducted on the Sunway TaihuLight supercomputer are shown in Fig.~\ref{Perf_SW}.
The horizontal resolution is the finest $0.25 \times 0.25$.
The model scales to 1.06 million cores (MPEs + CPEs).
The overall execution time is split into the computation time and the communication time for the MPI and the MPI\,+\,OpenACC implementationS, respectively.
As can be seen, the MPI\,+\,OpenACC acceleration achieves a 2.4x average and a maximum 4.1x speedups over the MPI implementation in terms of the total computation time.
The overhead of MPI communication is also reduced to a large extent owing to the loop reconstruction.
In Fig.~\ref{speedup_SW}, the speedup and efficiency of the scaling tests from 33,280 to 1,064,960 cores are exhibited.
Compared to 33,280 cores, the simulation of one million cores accomplishes 11.6x performance enhancement, obtaining 36.1\% parallel efficiency, which is comparable to that of the other platforms.
Considering that the Sunway TaihuLight was first released 5 years ago, and more specific and thorough optimizations for the next-generation \emph{Sunway protocol} are required, the absolute performance measured in SYPD has room for more improvements.
In sum, these results demonstrate the scalability of AGCM-3DLF on the heterogeneous supercomputer TaihuLight.

\section[]{Conclusion}
\label{sec:conclu}

In this work, we propose a highly scalable decomposed atmospheric general circulation model with a newly designed parallel finite-difference computing scheme,
i.e. the AGCM-3DLF, which scales well on different platforms at fine horizontal resolutions.
In comparison with the naive implementation, the 3D decomposition method in IAP-AGCM allows parallelism to be released in all three physical dimensions (longitudinal, latitudinal, and vertical), and significantly increases the parallelism and scalability of the model.
Additionally, the overall communication overhead is significantly decreased using the new 3D domain decomposition.
From the perspective of atmospheric sciences, the adoption of the new leap-format finite-difference computation scheme plays an equivalent part as the high latitude filtering modules. With the leap-format scheme, no extra filters are needed at high-latitudes ($|\varphi|>70\degree$). Thus, the overall communication overhead is significantly reduced and the load imbalance is relieved.
To integrate the 3d decomposition and the leap-format difference scheme, a novel shifting window communication algorithm for parallelization is introduced. 
And several optimizations are utilized for the efficiency of large-scale simulations, including decomposition prior strategy, refactoring of computation, 
aggregation of related variables, etc. 
To analyze the suitability for heterogeneous HPC platforms, a hybrid parallelization scheme and acceleration of stencil loops are adopted and the model is deployed on the heterogeneous many-core supercomputer Sunway TaihuLight.
Sufficient numerical tests and performance experiments for the complete AGCM model demonstrate the good efficiency and scalability of our new approach.
Experiments performed on the supercomputer CAS-Xiandao1 show that AGCM-3DLF scales from 4,096 processes to 196,608 processes,
i.e., the full system scale, and the simulation obtains a 13.1x speedup with the parallel efficiency of 27.2\%. And simulation results of AGCM-3DLF conducted on the Sunway TaihuLight supercomputer exhibit that AGCM-3DLF scales from 33,280 to 1,064,960 cores (MPE + CPEs),
obtaining a 13.1x speedup with the parallel efficiency of 27.2\%.

Our future work would further improve the performance of the atmospheric general circulation model to support the study of climate change.
And several efforts are ongoing to achieve better strong scalability and efficiency on the next-generation supercomputers.

\ifCLASSOPTIONcompsoc
  \section*{Acknowledgments}
\else
  \section*{Acknowledgment}
\fi

The authors would like to thank all the reviewers for their valuable comments. 
This work is supported by National Key R\&D Program of China under Grant No. 2016YFB0200803; the National Natural Science Foundation of China under Grant No. 61972376, No. 62072431, and No. 62032023; the Science Foundation of Beijing No. L182053.

\nolinenumbers
\ifCLASSOPTIONcaptionsoff
  \newpage
\fi



%
\bibliographystyle{IEEEtran}
\bibliography{IEEEabrv,AGCM-3DLF.bib}

\end{document}